\documentstyle[epsfig,graphicx,12pt]{article}
\begin{document}

\title{ Relationship of field-theory based single boson 
exchange potentials to current ones} 
\author{ A. AMGHAR$^1$, B.  DESPLANQUES$^{2}$
\thanks{{\it E-mail address:}  desplanq@isn.in2p3.fr} \\  
$^{1}$D\'epartement des Sciences Fondamentales, INHC, \\ 
  35000 Boumerdes, Algeria \\ 
$^{2}$Institut des Sciences Nucl\'eaires (UMR CNRS-IN2P3, UJF)\\ 
  F-38026 Grenoble Cedex, France }

\sloppy

\maketitle

\begin{abstract}
\small{
It is shown that field-theory based single boson exchange potentials 
cannot be identified to those of the Yukawa or Coulomb type that 
are currently inserted in the Schr\"odinger equation. The 
potential which is obtained rather corresponds to this current single boson 
exchange potential corrected for the probability that the system under 
consideration is in a two-body component, therefore missing contributions 
due to the interaction of these two bodies while bosons are exchanged. 
The role of these contributions, which involve at least 
two boson exchanges, is examined. The conditions that allow one to recover 
the usual single boson exchange potential are given. It is shown that 
the present results have some relation: i) to the failure of the 
Bethe-Salpeter equation in reproducing the Dirac or Klein-Gordon 
equations in the limit where one of the constituent has a large mass, 
ii) to the absence of corrections of relative 
order $\alpha \; {\rm log} \, \frac{1}{\alpha}$ to a full calculation of the 
binding energy in the case of neutral massless bosons  or iii) to 
large corrections of wave-functions calculated perturbatively in 
some light-front approaches.}
\end{abstract} 

\section{Introduction}  
It is well known that the Bethe-Salpeter equation \cite{BETH}, in 
both the ladder approximation and in 
the limit where one of the two interacting particles has a large 
mass, does not reproduce the Dirac or Klein-Gordon equations \cite{ITZY}. 
These are important benchmarks, as they accurately describe the 
interaction of charged particles with spin 1/2 (electrons, muons,...) 
or spin 0 (pions,...) in the Coulomb field produced by a nucleus. The 
remedy to the above failure resides in the contribution of second 
order crossed diagrams (see refs \cite {FRIA1, ITZY} for some review). 
At first sight, taking into account these diagrams is somewhat 
surprising as both the Bethe-Salpeter and the Dirac or Klein-Gordon 
equations imply the iteration of what looks like a single boson exchange, 
allowing one to sum up the contribution of ladder diagrams. There 
is no obvious  indication that these equations 
explicitly involve crossed diagrams. One can thus expect that the 
role of crossed dia\-grams in restoring the equivalence of the 
Bethe-Salpeter and Dirac or Klein-Gordon equations is partly 
fortuitous and results from some cancellation of contributions, 
which may not have a general character. 

It is also well known that the Bethe-Salpeter equation or the 
light-front approach, when applied to the Wick-Cutkosky model 
(two scalar particles exchan\-ging a massless spin 0 
particle \cite{WICK, CUTK}, leads to corrections 
of the binding energy with an unusual character \cite {FRIA1, FELD, ITZY}. 
In the non-relativistic limit, this model is equivalent to the Coulomb 
problem and the binding energy can be expressed in terms of 
the quantity, $m \, \alpha^{2}$, where $\alpha$ can be identified 
to the similar quantity used in QED ($\alpha$ is equal to the 
ratio of the coupling constant $g^{2}$, generally referred to in the model, 
with the quantity $4m^{2}.4\pi$). With these definitions, the corrections 
are of order $m \, \alpha{^3} \; {\rm log} \, \frac{1}{\alpha}$ (relative 
order $\alpha \; {\rm log} \, \frac{1}{\alpha}$ with respect to the binding 
ener\-gy). These ones have been determined theoretically \cite{FELD} 
and are present in some numerical calculations \cite{FRED, DESP2, CARB1}. 
It has been suggested that the difference with the Coulomb 
interaction case, where these corrections are known to be absent, 
would be due to the spin 1 nature of the photon \cite{FELD}.

A third point we would like to mention deals with the determination 
of a re\-lativistic wave function in the light-front approach for a 
realistic physical system. This represents a tremendous task, especially 
in the case of the NN system where the exchange of several mesons with 
different spins or isospins has to be consi\-de\-red. Present perturbative 
calculations using current deuteron wave functions as a zeroth 
order require a large renormalization of the first order wave 
function, by about $20\%$ \cite{FRED,CARB2}. This is much larger 
than the corrections due to incorporating relativistic kinematical 
effects, such as those due to factors $( \frac{m}{E} )^{1/2}$, 
which amount to a few percent \cite{PFEI}. 

This is also much larger than corrections due to renormalization 
effects, which are related to the meson in flight content of the 
deuteron state and have been found to be of the order 
of $3-4\%$ \cite{BONN, DESP1}. Furthermore, the use of this wave 
function calculated perturbatively has revealed bad features such 
as an unacceptable ratio of the deuteron asymptotic normalizations 
$A_{D}$ and $A_{S}$, $\frac{A_{D}}{A_{S}}$, or a deuteron quadrupole 
moment, both exceeding experiment by  $20-25\%$. These discrepancies 
are too large to be considered as conventional relativistic corrections 
of the order $(\frac{v}{c})^2$. They rather suggest that the comparison 
is biased by the absence of essential ingredients in the calculation, 
which should be first completed  to provide physically relevant results.

Obviously, as mentioned in \cite{CARB3}, one expects these 
difficulties to be removed when the problem is treated in 
its totality, implying the determination of an appropriate 
nucleon-nucleon interaction model, as made in the full (energy 
dependent) Bonn model \cite{BONN}. We nevertheless believe that 
it is important to understand the origin of the large renorma\-lization 
as well as of unexpected results in relation with the deuteron 
D-state. On the one hand, the difficulty in determining a 
nucleon-nucleon interaction model invites to rely as much as 
possible on what has already been achieved in the past. On the other, this 
understanding should avoid to blindly absorb in the parameters 
of a fitted single meson exchange potential contributions that 
are too specific to be treated in a so crude way. This remark 
is relevant for the $2\pi$-exchange part. Although it concerns 
a case which is less realistic than the NN one, a relatively 
large correction of the wave function was also found in the 
light-front approach applied to the Wick-Cutkosky model 
in the limit of a small coupling \cite{DESP2, CARB1}. This 
correction is located in the low momentum range of the wave 
function and, contrary to that one anticipated in \cite{KARM}, 
which rather occurs in the high momentum domain, it does not go 
to 0 with the momentum. For some part, it can be accounted for by a 
renormalization of the coupling constant, $ \alpha \rightarrow \alpha 
\,(1-\frac{2}{\pi}\alpha\,{\rm log} \, \frac{1}{\alpha})$. 

All the problems briefly sketched above have been raised 
in the framework of relativistic descriptions of a two-body 
system, such as the Bethe-Salpeter equation, or the light-front 
equation as used in \cite {FRED, CARB3, KARM}. What we would 
like to show here is that they are rather due to the field-theory 
treatment which underlies them and that they have a much more 
general character, with a close relation to dynamical effects in 
the field of the many body problem and the related derivation 
of effective degrees of freedom. Essentially, the effects under 
consideration rather have a static character and are different 
from usual relativistic effects that are velocity dependent and 
represent a higher order effect. 
To emphasize our point, we will essentially show  how the same 
problems occur and are solved in an approach which is mainly a 
non-relativistic one. While doing so, we are aware that the 
distinction is not straightforward. A field-theory treatment, 
which implies the creation and destruction of particles, is 
one of the ingredients that are required to fulfill the Lorentz 
covariance of some approaches like that followed in \cite{KARM,CARB3}. 
However, as evidenced by the very existence of relativistic 
quantum mechanics approaches \cite{KEIS}, this property can 
be ensured without relying on field theory. On the other hand, 
a consequence of the present work should be to provide a better 
zeroth order approximation for perturbative calculations using 
energy dependent interaction models and aiming to study relativistic 
effects. 

The plan of the paper is as follows. In the second section, starting 
from a field-theory based single boson exchange contribution to the 
two-body interaction, we remind how one can derive an energy independent 
effective interaction to be employed in a Schr\"odinger equation. 
The difference with the current single boson exchange potential 
of the Yukawa or Coulomb type is emphasized. The third section is 
devoted to the corrections that the above difference produces, 
namely some renormalization of the two-body wave function 
and  $\alpha^{3}\;{\rm log}\,\frac{1}{\alpha}$ terms in the binding energy 
in the case of an exchange of a massless (scalar) boson. In the 
fourth part, we consider the contribution due to including two 
boson exchanges, arising especially from the crossed diagram. It 
is shown how this allows one to recover current single boson exchange 
potentials in the case where the boson is unique (neutral) and couples 
predominantly in a scalar way with the constituent particles as 
in the Dirac or Klein Gordon equations. A simple explanation for 
the role played by this contribution is proposed. Some 
attention is given to the massless case, with regard to 
possible $m \, \alpha^{3} \; {\rm log}\,\frac{1}{\alpha} $ corrections to the 
binding energy, and to the nucleon-nucleon interaction case, 
with regard to the spin and isospin of the exchanged mesons. 
Some of the results presented here may be partly known in various 
domains of physics, but do not seem to be (or have been forgotten) in 
the domain of intermediate energy physics, at the border line of 
particle and nuclear physics. While they may not be quite new, we 
believe that reminding them here will make the paper more consistent.

\section{Contribution of the field-theory based single boson exchange to the 
current two-body effective interaction}

\begin{figure}[htb!]
\begin{center}
\mbox{ \epsfig{ file=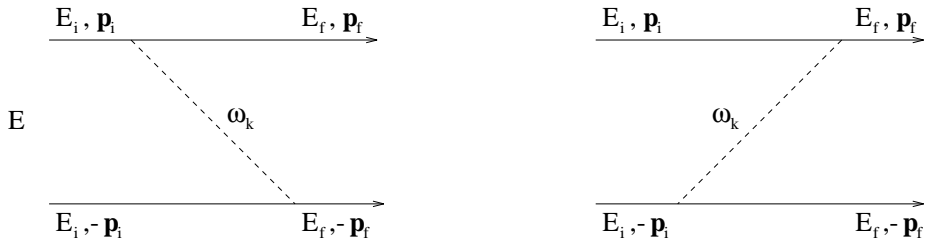, angle=270, width=13cm}}  
\end{center}

\caption{ \small {Time-ordered single boson exchange contributions to the 
two-body 
interaction with indication of the kinematics in the center of mass.}  }
\end{figure}  

Usual two-body interaction models suppose the instantaneous propagation 
of the exchanged bosons, which results in the conservation of the number 
of particles. In a field theory approach, the number of particles is not 
conserved and the two constituent particles may be accompanied by one, 
two,... bosons in flight (beside the excitation of anti-particles 
ignored in the present work). In this approach, which has a more 
fundamental basis \cite{BONN} and is much more ambitious, the lowest 
order interaction of two constituent particles, possibly bare, can 
be obtained by summing the contributions of the two time ordered 
diagrams shown in Fig. 1. In the center of mass system, it reads:
\begin{equation}
V_E(\vec{p}_i,\vec{p}_f)= 
\sum_{v}\frac{g^{2}_{v} \, O^{1}_{v} \, O^{2}_{v}}{2\omega^{v}_{k}}
\left( \frac{1}{E-\omega^{v}_{k}-E_{i}-E_{f}}
 +     \frac{1}{E-\omega^{v}_{k}-E_{i}-E_{f}} \right).
\end{equation}
In this equation, the summation on the index $v$ is carried over 
the different species of bosons, if any. The quantities $g_{v}$ 
and $O_{v}$ represent the coupling constant relative to the 
interaction of the boson with the constituent particles and its 
vertex function. This one may involve the spin, the isospin 
and some form factor. It also contains normalization factors 
relative to the constituent particles, $ \sqrt{\frac{m}{E_i}} $ 
and $ \sqrt{\frac{m}{E_f}} $, so that $g$ will be in what follows a 
dimensionless constant. The quantity, $\omega^{v}_{k},E_{i},E_{f}$ 
represent the on-mass shell energy of the exchanged boson and the 
constituent particles. They are defined as :
\begin{eqnarray}
\omega^{v}_{k} = \sqrt{\mu^{2}_{v}+\vec{k}^{2}},\nonumber \\
E_{i}= \sqrt{m^{2}+\vec{p}_i^2},\nonumber \\
E_{f}= \sqrt{m{^2}+\vec{p}_f^2},
\end{eqnarray}
where $\vec{k},\, \vec{p}_{i},\, \vec{p}_{f}$ represent the 3-momenta of the 
particles 
under consideration (in the center of mass for the two last ones) 
(see Fig. 1 for the kinematics). As to E, it represents the 
energy of the system we are interested in. Its presence in Eq. (2) 
is a characteristic feature of the coupling of the two-body component to 
components that are not explicitly specified with three or more bodies in the 
general case.

When the contribution of $V_E$ to the Born scattering amplitude 
is considered, the energy conservation relation, 
$E=2 \, E_{i}=2 \, E_{f}$, is fulfilled 
and it can then be checked that the two terms in Eq. (1) sum 
up to reproduce the standard Feynman expression for the 
propagator of a boson. Contrary to this one however, Eq. (1) 
provides a well determined procedure for extrapolating the 
interaction off-energy shell. The dependence of the 
interaction, $V_E$, on the energy $E$, which has the same 
origin as the one present in the full Bonn model of the $NN$ 
interaction \cite{BONN} (not to be confused with the other 
Bonn models denoted Q, R, QA, QB, QC, RA, RB \cite{BONN,MACH}), 
is at the same time a distinctive feature of a field-theory 
approach, but also a source of difficulties \cite{FRIA1}. Indeed, 
inserting $V_E$ in a Schr\"odinger equation leads to solutions 
that are not orthogonal in the case of states with different 
energies. This violates an important property expected 
from quantum mechanics, therefore suggesting that this way of 
proceeding is wrong or, at least, incomplete. While this 
incompleteness may be remedied by inserting contributions 
due to the many body components that the energy dependence 
of $V_E$ implies (these may be 2 constituent particles plus 1, 2,... 
bosons in flight), most often, the use of energy dependent 
potentials has been discarded, not without any reason \cite{FRIA2}.
Another way to remedy these difficulties is to derive 
from $V_E$, Eq. (1), an energy independent, but effective 
interaction. Many works along these lines have been performed 
in the literature \cite{FST,OKUB,  JOHN, SATO}. Here, we will proceed 
in a way which is perhaps more appropriate to emphasize the physics 
hidden behind the above energy dependence \cite{DESP1}. The 
interaction $V_E$, Eq. (1), is expanded in powers of $1/\omega_{k}$.
Starting from:
\begin{equation}
\frac{1}{  E -\omega_{k} -\frac{\vec{p}_i^2}{2m} -\frac{\vec{p}_f^2}{2m}  }=
-\frac{1}{\omega_{k}}
-\frac{ E-\frac{\vec{p}_i^2}{2m}-\frac{\vec{p}_f^2}{2m} }{ \omega^{2}_{k} 
}+\cdots ,
\end{equation}
where the energies of the total system and of the constituent 
particles are, from now on, replaced by the non-relativistic 
counterparts, one gets:
\begin{equation}
V_E = V_{0} - \frac{1}{2} \{ (E-\frac{\vec{p}^{2}}{m}),V_{1} \} \cdots.
\end{equation}
A different expansion around some energy, $E_0$, (renormalization point) 
could also be performed. More appropriate to the framework of the 
renormalization theory, it is less convenient here where a comparison with 
standard descriptions of two-body interactions is intended. The difference, 
which involves higher order terms in Eq. (3), is irrelevant for the present 
work, essentially devoted to a non-relativistic description of two-body systems.

The first term in the r.h.s. of Eq. (4), which stems from the 
similar one in Eq. (3), can be identified to a usual boson 
exchange interaction. In momentum space, it reads:
\begin{equation}
V_{0}(k)=-\sum_{v}\frac{g^{2}_{v} \, O^{1}_{v} \, 
O^{2}_{v}}{\mu^{2}_{v}+\vec{k}^{2}},
\end{equation}
In configuration space (for the simplest case where the vertex functions 
$O_{v}$ do not depend on momenta and where form factors are neglected), its 
expression is given by:
\begin{equation}
V_{0}(r)=-\sum_{v}g^{2}_{v} \, O^{1}_{v} \, O^{2}_{v} \;
     \frac{ e^{-\mu_{v}r} }{4\pi r},
\end{equation}
The quantity $V_{1}$ in Eq. (4), which is related to $V_E$ by the relation :
\begin{equation}
V_{1}(k) = -\frac{\partial V_E}{\partial E},\; ({\rm up} \; {\rm to} \; 
\frac{1}{\omega^{4}} \; {\rm terms}), 
\end{equation}
is given, in momentum space, by:
\begin{equation}
V_{1}(k) = \sum_{w}\frac{g^{2}_{w} \, 
            O^{1}_{w} \, O^{2}_{w}}{(\omega^{w}_{k})^{3}},
\end{equation}
The corresponding expression in configuration space, in the same approximation 
as for Eq. (6), reads:
\begin{equation}
V_{1}(r) = \sum _{w}g^{2}_{w} \, O^{1}_{w} \, O^{2}_{w}\frac{1}{2\pi^{2}}
  \int\frac{dk k^{2}j_{0}(kr)}{(\omega^{w}_{k})^{3}},
\end{equation}
With the expansion of $V_E$ we made together with the assumption of 
local vertex functions, the dimensionless quantity $V_{1}(r)$ 
given by Eq. (9) is local. This feature greatly facilitates the 
interpretation of its role but will not be necessarily used 
throughout this paper. In the case of 
massless bosons, like in QED or in the Wick-Cutkosky model, the 
integral in Eq. (9) diverges logarithmically and some slight 
modification of the expansion given by Eq. (3) should be 
performed (see Sect. 3.2).

With the above definitions, the equation that the wave function 
describing the two-body components, $ \psi(r)$, has to satisfy reads:
\begin{equation}
\left(V_{0}(r) + \frac{1}{2} \{ \frac{\vec{p}^{2}}{m},(1+V_{1}(r)) \} - 
E(1+V_{1}(r))\right)\psi(r)=0.
\end{equation}
By multiplying on the left by $(1+V_{1}(r))^{-1/2}$ and making 
the substitution
\begin{equation}
\psi(r)=(1+V_{1}(r))^{-1/2} \, \phi(r),
\end{equation}
one gets an equation that can now be identified to a Schr\"ošdinger equation:
\begin{equation}
\left( V(r)+\frac{\vec{p}^{2}}{m}-E \right) \phi(r)=0,
\end{equation}
where
\begin{equation}
V(r) = (1+V_{1}(r))^{-1/2} \; V_{0}(r) \; (1+V_{1}(r))^{-1/2} + \dots.
\end{equation}
The dots in Eq. (13) represent a higher order contribution in both $V_{1}(r)$  
and $ \frac{1}{m} $, whose expression, in the limit of a local operator, is 
given by:
\begin{displaymath}
-\frac{1}{4m} \frac{(\vec{\nabla}V_{1}(r))^2}{(1+V_{1}(r))^2}.
\end{displaymath}
Being of the order of relativistic contributions neglected here (Z-diagrams for 
instance), it is not considered in the following. The derived potential, $V(r)$, 
can thus be approximated by:
\begin{equation}
V(r)= V_{0}(r)-\frac{1}{2} \{ V_{1}(r),V_{0}(r) \}+\cdots.
\end{equation}
This expansion may be useful for a discussion of corrections order by order. 
The full expression (13) is nevertheless more appropriate in the case where 
an expansion in terms of  $V_{1}$ is not legitimate.

At the lowest order in $V_{1}$, the 
correction to the potential $V_{0}$ reads in momentum space:
\begin{eqnarray}
\Delta V(\vec{p}_{i},\vec{p}_{f}) = \frac{1}{2} \int \frac{d \vec{p}}{(2\pi)^3} 
\;    \left( \sum_{w}\frac{g^{2}_{w} \, O^{1}_{w} \, 
O^{2}_{w}}{(\mu^{2}_{w}+(\vec{p}_{\,i}-\vec{p})^{2})^{\frac{3}{2}}} \;
\sum_{v}\frac{g^{2}_{v} \, O^{1}_{v} \, 
O^{2}_{v}}{\mu^{2}_{v}+(\vec{p}_{\,f}-\vec{p})^{2}} \right. 
\;\;\;\;\;\;\;\;\; \nonumber \\
 \left. +\sum_{v}\frac{g^{2}_{v} \, O^{1}_{v} \, 
O^{2}_{v}}{\mu^{2}_{v}+(\vec{p}_{\,i}-\vec{p})^{2}} \;
\sum_{w}\frac{g^{2}_{w} \, O^{1}_{w} \, 
O^{2}_{w}}{(\mu^{2}_{w}+(\vec{p}_{\,f}-\vec{p})^{2})^{\frac{3}{2}}} \right).
\end{eqnarray}
After making the change $w \leftrightarrow v $ in the second term and using 
quantities  $\omega_i^w$ and $\omega_f^v$ defined analogously to Eq. (2):
\begin{equation}
\omega_i^w=(\mu^{2}_{w}+(\vec{p}_{\,i}-\vec{p})^{2})^{\frac{1}{2}},
\;\;\;\;\;\;\;\;\;\;
\omega_f^v=(\mu^{2}_{v}+(\vec{p}_{\,f}-\vec{p})^{2})^{\frac{1}{2}},
\end{equation}
the above expression may be written in a simpler form:
\begin{equation}
\Delta V(\vec{p}_{i},\vec{p}_{f}) = \frac{1}{2} \int \frac{d \vec{p}}{(2\pi)^3} 
\;  \sum_{w,v} \left( \frac{g^{2}_{w} \,g^{2}_{v} \; O^{1}_{w} \,O^{2}_{w} \,  
O^{1}_{v} \,O^{2}_{v}}{\omega_i^{w^2} \omega_f^{v^2} }
 \; ( \frac{1}{\omega_i^w} +  \frac{1}{\omega_f^v} )  \right).
\end{equation}

This correction, which contains twice the coupling $g^2$ 
appears as a second order boson exchange contribution and, for a 
unique spinless scalar boson, tends to have a repulsive character. 
For consistency, it should be considered with contributions of 
the same order  (to be discussed in Sect. 4). The appearance 
of second order contributions in Eq. (13) may look surprising 
as the starting point given by the field-theory motivated 
interaction, Eq. (1), was a first order one. It originates from 
the off-energy shell behavior of this interaction, which, after 
elimination, transforms into a higher order correction. This 
currently occurs when changing the degrees of freedom used in 
describing some system to effective (or dressed) ones (see \cite{AMGH} for 
another example in 
relation with the NN interaction). Notice that $V_{0}(r)$ and $V_{1}(r)$ in 
Eq. (14), or factors $O^{1}_{w} \, O^{2}_{w}$ and $O^{1}_{v} \, 
O^{2}_{v}$ in Eq. (15), may not always commute. This respectively 
occurs for momentum dependent (non local) interactions or spin-isospin 
dependent ones.  In the following, we will sometimes disregard 
these cases, essentially for simplifying the discussion and 
providing a more  sensible presentation of the role of corrections 
associated with $V_{1}(r)$. 

\begin{figure}[htb!]
\begin{center}
\mbox{ \epsfig{ file=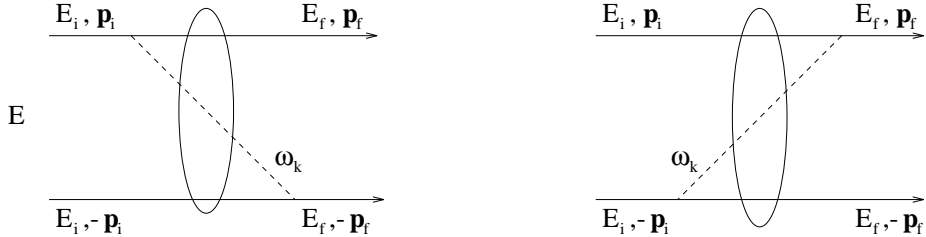, angle=270, width=13cm}}
\end{center}
\caption{ \small{Time-ordered boson exchange contributions to the two 
body interaction,  including some interaction between the two 
constituents (symbolized by the oval).} }
\end{figure}  

The operator, $V_{1}(r)$  ($V_{1}(r)/(1+V_{1}(r))$ more exactly) 
has a simple interpretation. It represents the probability that 
the system be in a component where the two constituent particles 
are accompanied by a boson in flight. The interaction $V(r)$ in 
Eq. (12) therefore corresponds to the interaction due to the component 
with two constituent particles only. Consistently with  contributions 
of diagrams of Fig. 1 retained until now, it does not contain any 
interaction of these particles while a boson is in flight (see 
Fig. 2 for a graphical representation). As to the reference to 
dressed particles, it makes it clear that higher order corrections 
should appear in the effective potential, as mentioned above. 
These corrections have some relationship to the renormalization 
presented in a more abstract way in \cite{FST,OKUB,CHEM,GARI,HYUG,SATO}. 
Let's also mention the consistency of: i) the above interpretation 
of $V_{1}(r)$, ii) the general  definition of the normalization 
for an energy dependent potential like that given by 
Eq. (10) and iii) the relation given by Eq. (11): 
\begin{equation}
N = \int d\vec{r} \; \psi^2(r) (1-\frac{\partial V_E}{\partial E})   = \int  
d\vec{r} \; \psi^2(r) 
(1+V_{1}(r)) = \int d\vec{r} \; \phi^2(r),
\end{equation}  
Not surprisingly, there is a relationship to the renormalization in 
the field of the many problem. The quantity, $(1+V_{1}(r))^{-1}$, 
is quite similar to the spatially dependent spectroscopic factor, 
which is introduced in the domain of mean field approximations to 
account for the effect of degrees of freedom (nuclear correlations) 
that have been eliminated (see ref. \cite{MAHA} for a review). 
Its effect, or what accounts for it to be more correct, 
turned out to be essential to preserve the orthogonality of many 
body wave functions when incorporating contributions due to the 
same degrees of freedom \cite{NOGU}.

At this point, it thus appears that the interaction $V(r)$, which is 
obtained from a field-theory based single boson exchange interaction 
does not identify to the single boson exchange potential $V_{0}(r)$ 
which, most often, is referred to and is used in the Schr\"ošdinger 
equation, or an equivalent one.

In the case of the strong $NN$ interaction, the above problem may 
not be a serious source of concern. Its short range part has some 
phenomenological character and through the fit of the potential 
parameters to experimental data, it may be possible to account 
in some part for the correction to $V_{0}(r)$ approximately given 
by $\frac{1}{2}\{V_{1}(r),V_{0}(r)\}$, which also has a short range character. 
Only the longest range part of this term which corresponds to two pion 
exchange and has a complicated spin isospin structure, may not be well 
reproduced.

In the case of the Coulomb potential, where the fundamental interaction 
is known, there is no uncertainty similar to that affecting the short 
range part of the $NN$ interaction. As contributions provided by the term, 
$\frac{1}{2}\{V_{1}(r),V_{0}(r)\}$, are known to be absent from the energy 
spectrum of atomic bound states involving electrons or other charged particles, 
the derivation of the potential given in this section has to be 
completed in any case. This will be done in Sect. 4, while contributions 
will be estimated in the next section.

It is worthwhile to mention that 
removing the energy dependence of the potential $V_E$ is 
quite similar to the Foldy-Wouthuysen transformation for 
the Dirac equation. Like this one, it is not merely a 
mathematical operation. From a state which is a superposition 
of a two-body and a two-body plus one boson in flight 
components, which respectively play the role of the large and 
small components in the Dirac spinors, it allows one to 
build a state with a unique two-body (dressed) component. 
This one represents a frozen coherent superposition of the 
bare components, with the consequence that the constituent 
particles have now acquired in the transformation an effective 
character. The change in the nature of the constituents is 
better seen by working in Fock space \cite{DESP3} 
(see also the appendix for an example). Quite similarly to the 
Foldy-Wouthuysen transformation, the deve\-lopments presented in 
the present section make sense around some given energy (states close to 
threshold in the present work). On the one hand, the choice of 
this energy, which plays the role of the renormalization 
point in the renormalization theory, could be optimized. On the 
other hand, by removing higher order energy dependent terms 
in the expansion given by Eq. (3), one can enlarge the energy 
domain at will, as far as one does not cross the inelasticity 
threshold where the frozen degrees of freedom may show up. 
While doing so however, the effective (or renormalized) interaction 
$V(r)$ in Eq. (13) acquires a non local character whose 
role will steadily increases when approaching this threshold.

The contribution considered here may have been considered as a 
relativistic effect and, as such, as a new contribution to be 
accounted for. It implies a non-instantaneous propagation of the 
exchanged boson, but this feature is also pertinent to a two-body system 
that, under the effect of some interaction, undergoes a transition 
from a low momentum component to a high momentum one and back, or 
a two constituents system (NN for instance) that undergoes a transition to a 
component involving their internal excitation ($ \Delta \Delta $)  
and back. These processes, which can give rise to an effective 
interaction, are typically part of a non-relativistic description 
of a two-body system. Contrary to current relativistic effects 
that are of the order $(v/c)^2$, they have a static character and 
don't vanish when the mass of the constituents goes to infinity.

The current relativistic effects and the contribution of interest 
here can be related to different contributions in the expression 
which, for instance, allows one to calculate some amplitude from an 
interaction acting on an input amplitude:
\begin{equation}
I= \int d\vec{q} \int dq_0 \frac{f(q_0,....)}{ 
((\frac{1}{2}P_0 +q_0)^2 -m^2-....)((\frac{1}{2}P_0 -q_0)^2 -m^2-....)
} \, \frac{1}{\omega^2 - q_0^2}.
\end{equation}
In this expression, where one recognizes the propagators of 
the constituent particles as well as the propagator of the 
exchanged boson, $q_0$ and $P_0$ represent the time component of 
the 4-momentum carried by the boson and the system under consideration. 
Contributions  to the integral come from the poles in the integrand.
The poles arising from the constituent 
propagators generally imply a small value of $q_0$, determined 
by the energy transfer of a constituent in the initial state to a 
constituent in the final state. They give rise to the above mentioned 
relativistic effects of the order $ \omega^{-4}$. The pole due to 
the boson propagator implies a value of $q_0$ that by no means 
is small. It is responsible for the effect mainly considered here, 
of the order $ \omega^{-3}$. While it is conceivable that the first effect 
disappears in the limit of an instantaneous propagation of the boson, 
which supposes that $q_0$ can be neglected in the boson propagator 
in Eq. (19), this is difficult to imagine for the second one.
It is also noticed that the two contributions are on a  quite different 
footing. The second one disappears when the boson propagator in Eq. (19) 
is replaced by the propagator $(\mu^2 - q^2 + \frac{(P.q)^2}{P^2})$, 
which is equivalent for a physical process (Born amplitude). On the 
other hand, as it can be seen from the expression of the denominator 
in Eq. (3), the effect which is considered here is of the 
order $\mu/m$, i.e. a recoil effect. 
\section{Estimates of the contributions 
   due to the term $\frac{1}{2}\{V_{1},V_{0}\}$}
   
We present in this section results of perturbative calculations 
of two-body wave functions with and without including in the 
Schr\"odinger equation the contribution due to the term 
$\frac{1}{2}\{V_{1},V_{0}\}$. At the same time, we will discuss the 
relationship of the wave functions so obtained with those calculated 
from an energy dependent interaction. The aim is to show that a large part of 
the physics which is at the origin of this energy dependence can be accounted 
for in an energy independent scheme (where the underlying degrees of freedom 
have acquired an effective character). In practice, this will be done 
for the NN interaction case and the Wick-Cutkosky model. 

The wave functions of interest are given in momentum space by:
\begin{eqnarray}
\tilde{\phi_0}(\vec{p})=-\frac{m}{p^2+\kappa^2} \int \frac{ d\vec{p\,}' 
}{(2\pi)^3}  \; 
V_0(\vec{p},\vec{p\,}') \; \phi_0(\vec{p\,}') , \\ 
\tilde{\phi}(\vec{p})=-\frac{m}{p^2+\kappa^2} \int \frac{ d\vec{p\,}' 
}{(2\pi)^3}  \;
V(\vec{p},\vec{p\,}') \; \phi_0(\vec{p\,}').
\end{eqnarray}
where  $V(\vec{p},\vec{p\,}')$ differs from $V_0(\vec{p},\vec{p\,}')$ 
by the correction given by Eq. (15). The quantity $\kappa^2$ is 
related to the binding energy by the relation $|E|=\frac{\kappa^2}{m} $. 
By comparing the results, we can estimate the effect of this 
correction. In principle, if $\phi_0(\vec{p})$ is chosen as the 
solution of the Schr\"odinger equation with the potential 
$V_0(\vec{p},\vec{p\,}')$, $\tilde{\phi}_0(\vec{p})$ is equal to 
$\phi_0(\vec{p})$ (strictly speaking,  this is not anymore a 
perturbative calculation but, rather, a consistency check). In 
some cases however, some terms in the original potential model 
that was used to calculate $\phi_0(\vec{p})$, may be neglected. 
It is then more appropriate to look at the effect 
of the correction by comparing $\tilde{\phi}(\vec{p})$ and 
$\tilde{\phi}_0(\vec{p})$, rather than $\tilde{\phi}(\vec{p})$ and 
$\phi_0(\vec{p})$, hence the introduction of $\tilde{\phi}_0(\vec{p})$.
On the other hand, while a schematic description of effects under 
consideration here has been given in the previous section, in practice 
some of the calculations presented in the present section, with the Bonn 
models, have been performed with the full complexity of the interaction. 
This involves in particular the non-locality due to describing spin 1/2 
particles by Dirac spinors and normalization factors appropriate to the 
definition of the potential V (together with the corresponding equation). 
For these cases, the numerical accuracy has been pushed to the point 
where one can reproduce the numerical deuteron wave function at a 
level where the discrepancy cannot be seen in the figures 
presented below.

Other quantities of interest correspond to perturbative 
solutions of the Schr\"odinger equation where the potential 
$V_0(\vec{p},\vec{p\,}')$ is replaced by the energy dependent 
potential given by Eq. (1):
\begin{equation}
\tilde{\psi}(\vec{p})=-\frac{m}{p^2+\kappa^2} \int \frac{ d\vec{p\,}' 
}{(2\pi)^3}  \; 
V_E(\vec{p},\vec{p\,}') \; \psi_0(\vec{p\,}').
\end{equation}
Making an expansion up to the first order term in $V_1$, the above 
equation may also read: 
\begin{eqnarray}
\tilde{\psi}(\vec{p}) \simeq -\frac{m}{p^2+\kappa^2} \int \frac{ d\vec{p\,}' 
}{(2\pi)^3}  
\;\;\;\;\;\;\;\;\;\;\;\;\;\;\;\;\;\;\;\;\;\;\;\;\;\;\;\;\;\;\;\;\;\;\;\;\;\;\;\;
\;\;\;\;\;\;\;\;\;\;\;\;\;\;\;\;\;\;\;  \nonumber \\ 
\left( V_0(\vec{p},\vec{p\,}')+ 
\frac{1}{2} ( \frac{p^2+\kappa^2}{m} \, V_1(\vec{p},\vec{p\,}') + 
V_1(\vec{p},\vec{p\,}') 
\, \frac{p'^2+\kappa^2}{m} ) \right) \psi_0(\vec{p\,}') \nonumber \\
\simeq -\frac{m}{p^2+\kappa^2} \int \frac{ d\vec{p\,}' }{(2\pi)^3}  \left( 
V_0(\vec{p},\vec{p\,}')+ 
\frac{1}{2}  
V_1(\vec{p},\vec{p\,}')  \, \frac{p'^2+\kappa^2}{m} \right) \psi_0(\vec{p\,}') 
\nonumber \\ 
-\frac{1}{2}\int \frac{ d\vec{p\,}' }{(2\pi)^3}  \; 
V_1(\vec{p},\vec{p\,}') \; \psi_0(\vec{p\,}'). 
\;\;\;\;\;\;\;\;\;\;\;\;\;\;\;\;\;\;\;\;\;\;\;\;\;\;\;\;\;\;\;  
\end{eqnarray} 
Equation (22) (together with Eq. (23)) differs from Eq. (21) by the 
fact that the first one involves the full energy dependent 
interaction $V_E$, while the second one involves the corresponding 
energy independent potential. It may be of interest to compare 
the outputs of these equations. It is remembered that  
$\phi_0(\vec{p})$ and $\psi_0(\vec{p})$ differ by an operatorial 
factor, $(1+V_1)^{\frac{1}{2}}$, see Eq. (11).  Two cases are 
to be considered: \\ 
i) The first one assumes that  $\psi_0$ (and not $\phi_0$) is solution of 
the Schr\"odinger equation with the potential $V_0$. One thus gets:
\begin{equation}
\tilde{\psi}= \psi_0 +\frac{1}{2} \, \frac{m}{p^2+\kappa^2} \int \frac{ 
d\vec{p\,}' 
}{(2\pi)^3} \, V_1 \, V_0 \, \psi_0 
-\frac{1}{2} \int \frac{ d\vec{p\,}' }{(2\pi)^3} V_1 \, \psi_0.
\end{equation}
This is formally close to the choice made in 
refs. \cite{CARB2, FRED}. \\ ii) The second case assumes that $\phi_0(\vec{p})$  
is solution of the Schr\"odinger equation with the potential $V_0$. 
According to Eq. (11), the zeroth order wave function to be 
inserted in Eq. (23) is related to the above one by the relation 
$\psi_0=(1+V_1)^{-\frac{1}{2}}\phi_0$. The wave 
function, $\tilde{\psi}(\vec{p}) $, then reads:
\begin{eqnarray}
\tilde{\psi}(\vec{p}) \simeq -\frac{m}{p^2+\kappa^2} \int \frac{ d\vec{p\,}' 
}{(2\pi)^3} 
\left( V_0(\vec{p},\vec{p\,}')-\frac{1}{2} \{ V_0 , V_1 \} \right) \; 
\phi_0(\vec{p\,}')  \nonumber \\        -\frac{1}{2}\int 
\frac{ d\vec{p\,}' }{(2\pi)^3}  V_1(\vec{p},\vec{p\,}') \; \psi_0(\vec{p\,}') 
\;\;\;\;\;\;\;\;\;\;\;\;\;\;\;\;\;\;\;\;\;\;\;\;\;\; \nonumber\\
\simeq (1-\frac{1}{2}V_1) \; \tilde{\phi}(\vec{p}) . 
\;\;\;\;\;\;\;\;\;\;\;\;\;\;\;\;\;\;\;\;\;\;\;\;\;\;\;\;\;\;\;\;\;\;\;\;\;\;\;\;
\;\;\;\;\;\;\;\;\;\;\;\;\;
\end{eqnarray}
This relation is in agreement with  Eq. (11), which relates the 
wave functions of the energy dependent and energy independent 
schemes, $\psi \simeq 
(1+V_1)^{-\frac{1}{2}}\phi \simeq (1-\frac{1}{2}V_1) \, \phi $. 

We now consider successively the Nucleon Nucleon interaction and the Wick 
Cutkosky (or QED) models.

\subsection{Nucleon Nucleon case}
We estimated corrections to the wave function arising from the 
term $\frac{1}{2} \{ V_0 , V_1 \}$, starting from a zeroth order 
wave function given by the Paris model \cite{PARI}. This was done 
perturbatively, using expressions given in the introduction of 
this section. In a first step, $V_0$ was approximated by a sum 
of a few well known meson exchanges $(\pi,\rho,\omega,\sigma)$, whose 
contributions in the standard non-relativistic limit were retained. 
The various parameters were chosen so that to reproduce the deuteron 
wave function of the model at large distances and as accurately 
as possible at short distances. Results obtained in this way have 
shown that the deuteron S wave function was strongly reduced, more than 
obtained in refs. \cite{CARB2,FRED}. The D wave was less changed, resulting 
in a relative enhancement of the D wave with respect to the S wave, 
qualitatively in agreement with the findings of these works. 
Due to the large effects, and to avoid to be misled by a hidden 
bias, we turned to a more complete calculation, with $V_0$ taken 
as identical to the potential used to calculate the deuteron wave 
function. In this order, we use meson exchange mo\-dels which, contrary 
to the Paris model, allow one to easily derive the operator, $V_1$. 
The choice of the Bonn  Q model will make possible a comparison with 
earlier work \cite{CARB2}. 
As this model provides a wrong prediction of the mixing angle, $\epsilon_1$, 
for energies beyond 100 MeV, we also use the Bonn QB model which 
provides a correct prediction with this respect and is generally 
considered as a quite reasonable model.

\begin{figure}[htb!]
\begin{center}
\mbox{ \epsfig{file=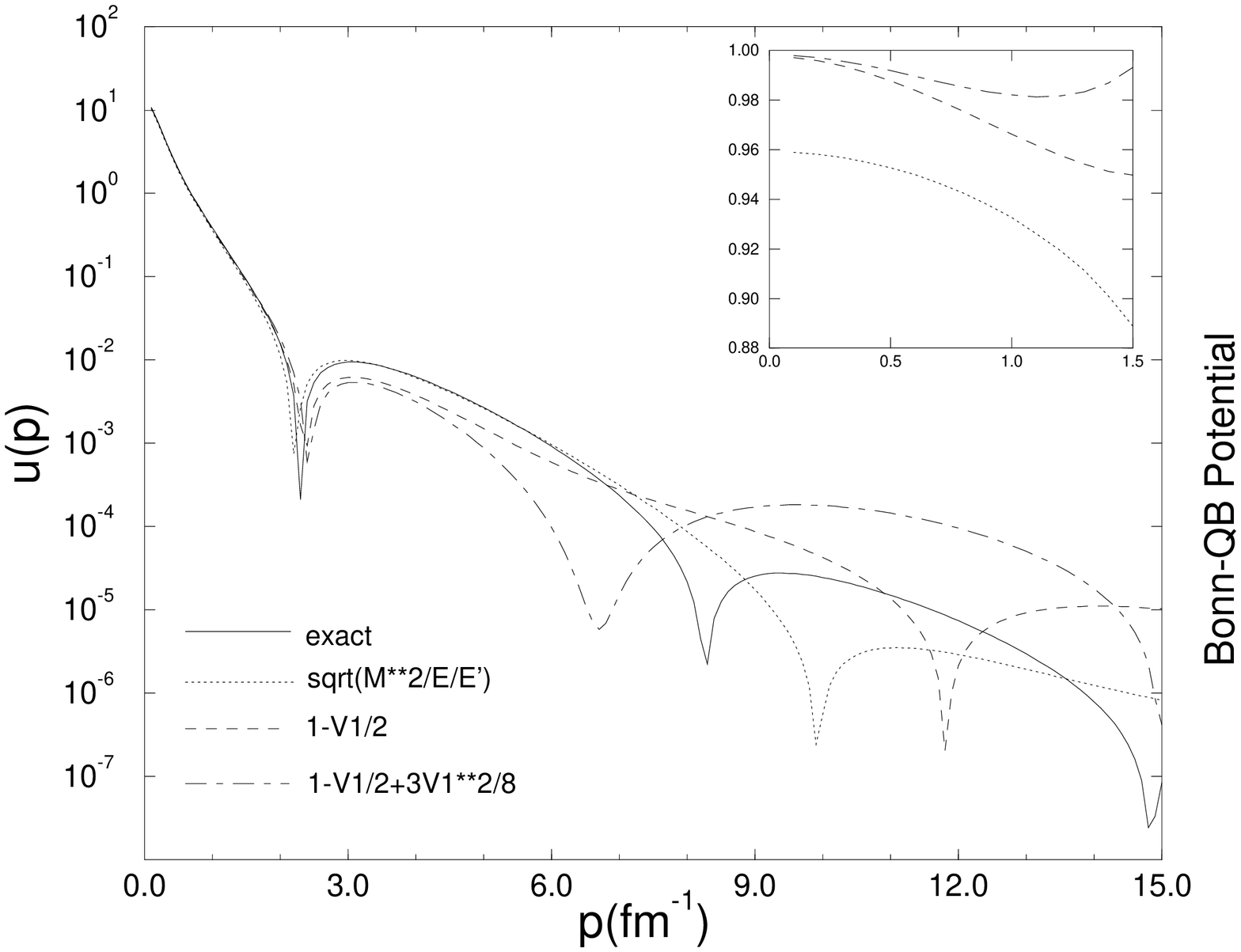, angle=270, width=10cm}  }
\mbox{ \epsfig{file=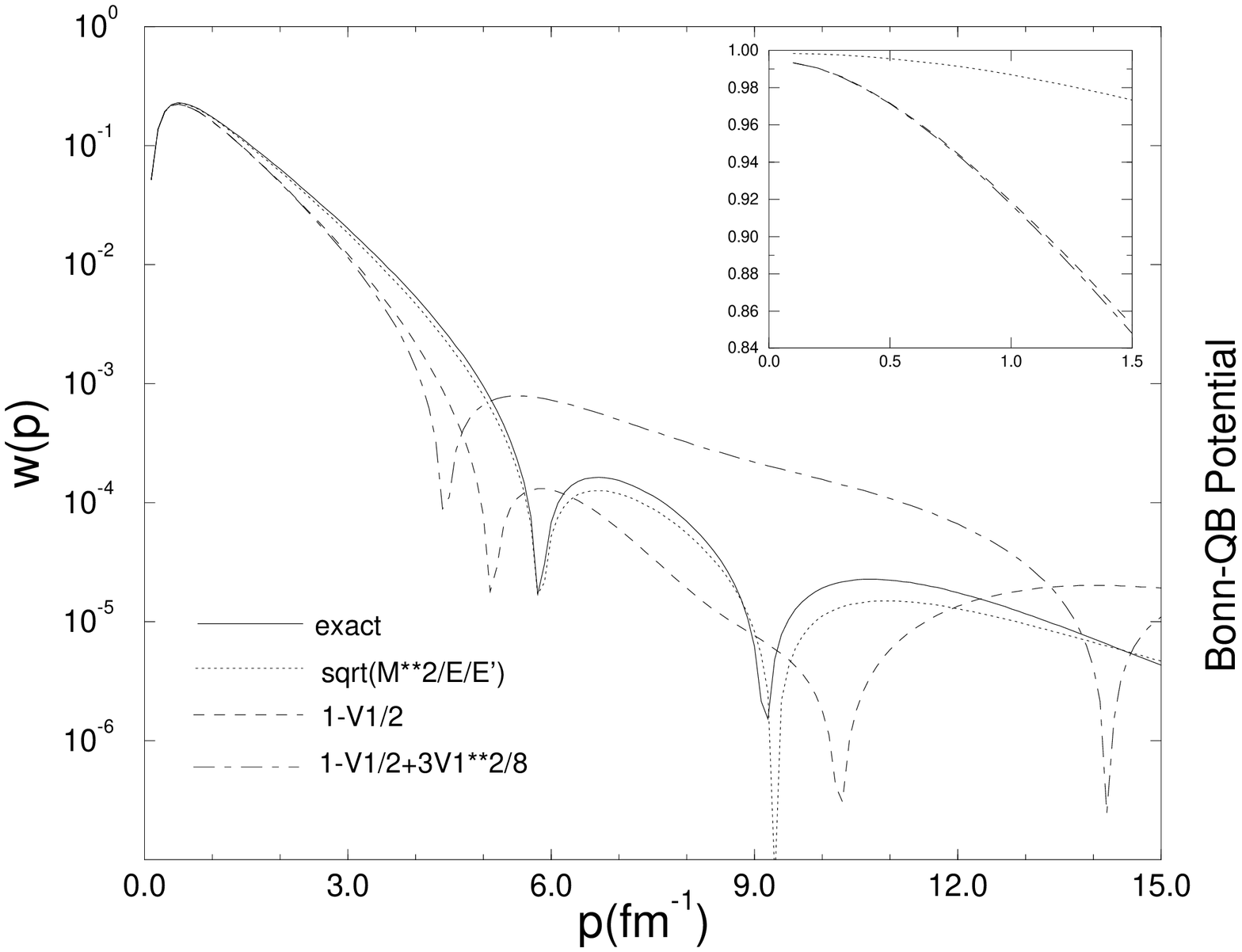, angle=270, width=10cm} }
\vspace{-0.5cm}
\end{center}      
\caption{\small{Sensitivity to relativistic kinematics and underlying schemes 
(dependence or independence of the total energy). The input wave function 
(continuous line) and the potential are those of the Bonn-QB model. 
The S and D components are presented respectively in the top and 
bottom parts of the figure. To better emphasize the small momentum behavior, 
we provide in the insets the ratio of the various calculations to 
the exact ones.}}
\end{figure}  

\begin{figure}[htb!]
\begin{center}
\mbox{  \epsfig{file=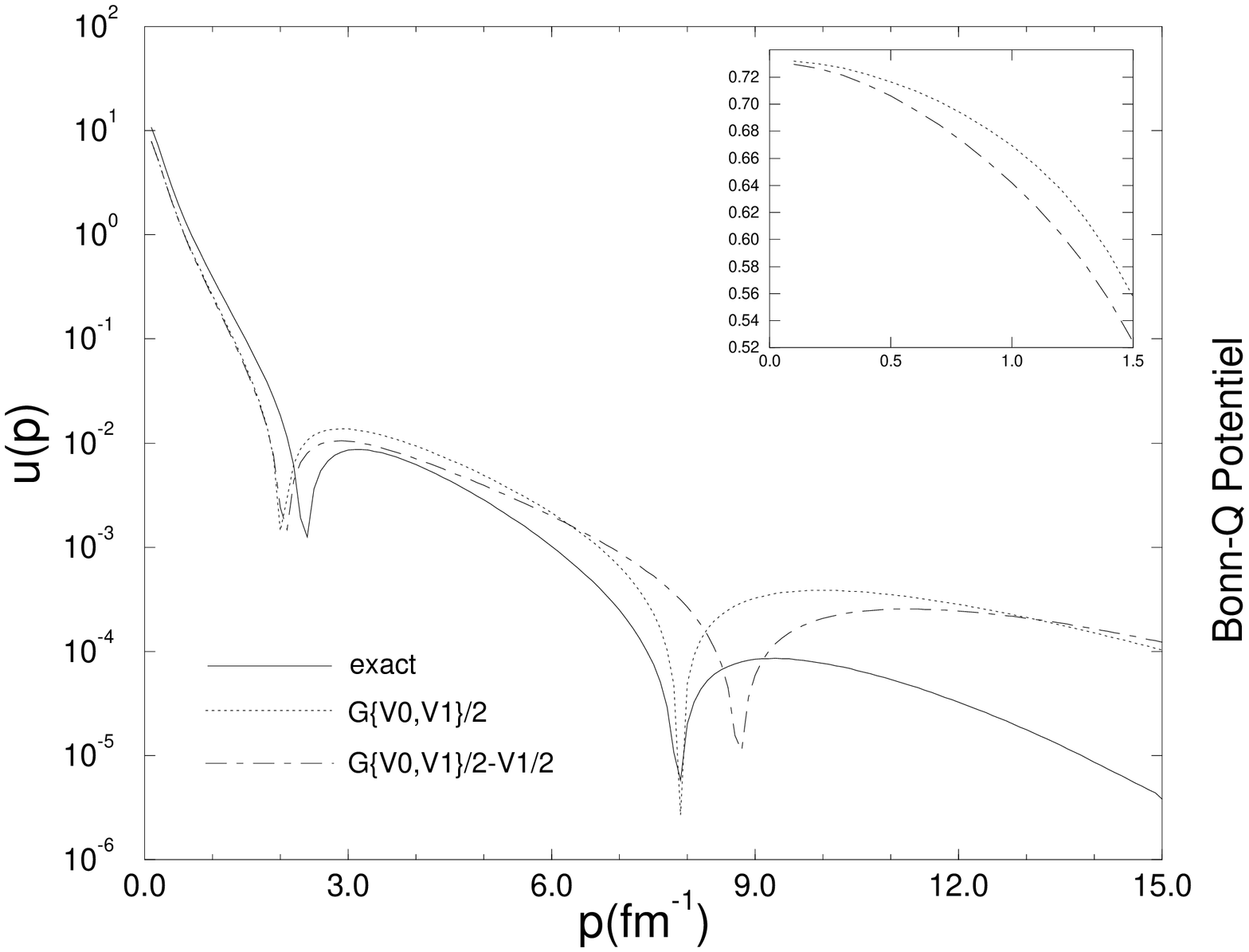, angle=270, width=10cm} }
\mbox{ \epsfig{file=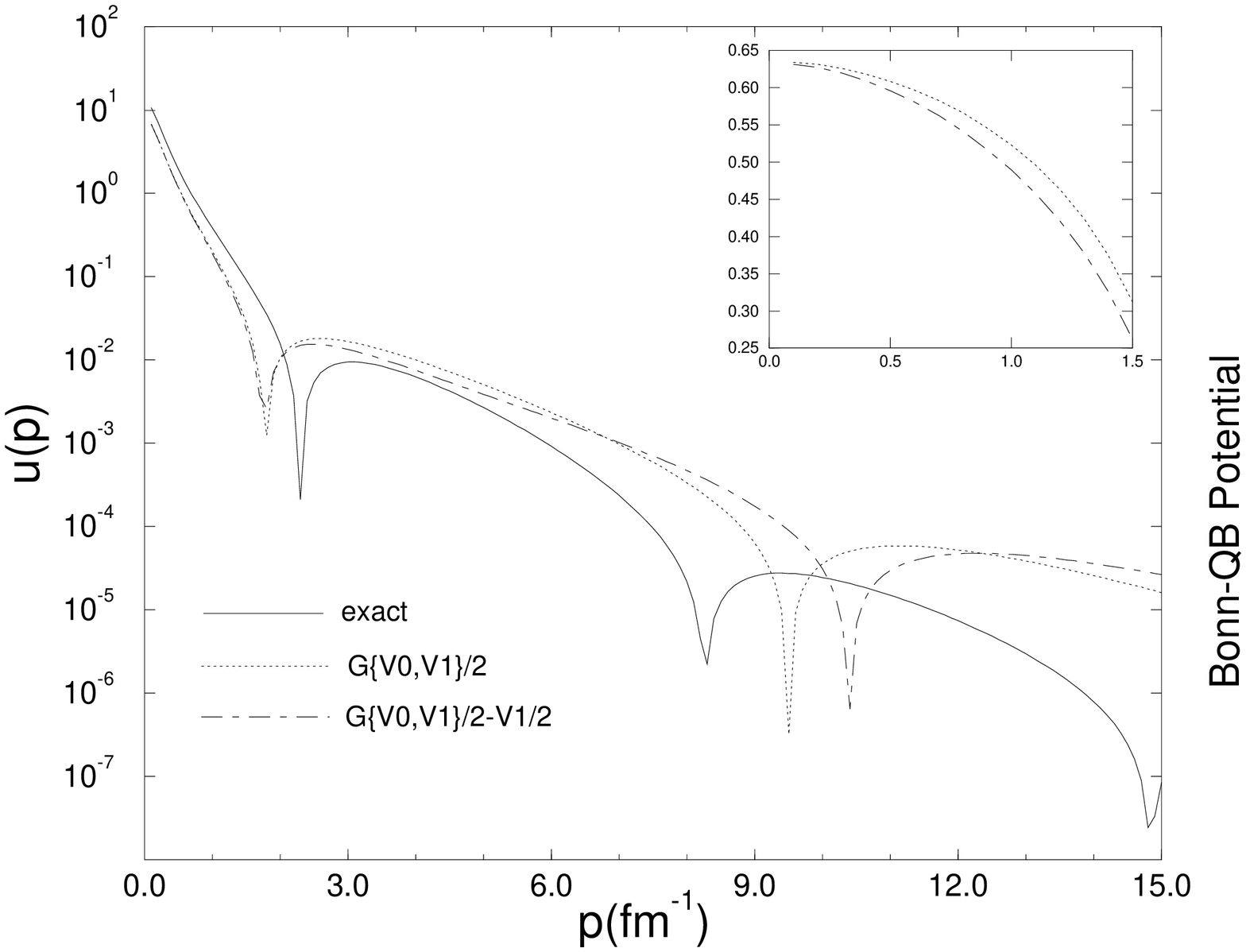, angle=270, width=10cm}  }
\vspace{-0.5cm}
\end{center}        
\caption{\small{Effect  of the  $\frac{1}{2}\{V_{1},V_{0}\}$  term, 
Eq. (14), on the calculation of the S component of the deuteron 
wave function. 
The input wave function (continuous line) and the potential are 
those of the Bonn-Q and Bonn-QB models (respectively upper and 
lower parts of the figure). The dotted and  dash-dotted lines correspond 
to outputs of the energy independent and energy dependent 
schemes respectively. The low momentum results are shown in the insets as  a 
ratio to the exact results.}}
\end{figure}

In Fig. 3, we first show the effect of introducing somewhat arbitrarily 
extra normalization factors $\sqrt{ \frac{m}{E} } \, \sqrt{ \frac{m}{E'} }$ 
in the integrand of Eq. (20). Calculations are performed with the 
Bonn QB model for both the deuteron S and D waves.  This can provide 
an estimate of the size of typical relativistic effects. The largest 
part, 20\% in the range $p=1 \, {\rm GeV}/c$ \cite{PFEI}, is given by 
the factor, $\sqrt{ \frac{m}{E} }$, which can be factorized out 
in Eq. (20). As the presence or absence of such a factor is dictated by 
unitarity conditions and is accounted for in many nucleon nucleon 
interaction models, the effect may not be a real one however. Some 
effect is also due to the other factor, $\sqrt{ \frac{m}{E'} }$, 
whose relevance was mentioned in \cite{CARB3}. 
It gives rise to a decrease of the S wave of 4\% at small momenta, 
which has the size of corrections generally expected from relativistic 
effects involving the internal structure, and of the order of 10-20\% 
in the range $p=1 \, {\rm GeV}/c$. The effect of 
the factor, $(1+V_1)^{-\frac{1}{2}}$, relating the wave functions 
in the energy dependent and energy independent schemes, is also 
shown in Fig. 3. Due to the difficulty in performing a full calculation, 
which has to do with the non-locality of the operator, we can only display 
the effect due to first and second order corrections in $V_1$. 
This however has the advantage of providing an estimate of some 
contributions neglected here. As expected, wave functions at small 
momenta are not modified. At intermediate momenta, some decrease 
is observed for both the S and D waves, which is compensated at 
the level of the norm by the contribution due to the meson in flight 
content of the order of 3-4\%. The results roughly agree with  
those obtained in ref. \cite{DESP1}, which were calculated exactly 
but with a localized form of $V_1$, or with that obtained for 
the full Bonn potential \cite{BONN}. Altogether, 
results presented in Fig. 3 show that neither relativistic effects, 
nor those dealing with the relation of the wave functions of the 
energy dependent and energy independent schemes can explain a 
large change of the deuteron S and D waves at small momenta 
(and equivalently at large distances).

\begin{figure}[htb!]
\begin{center}
\mbox{     \epsfig{file=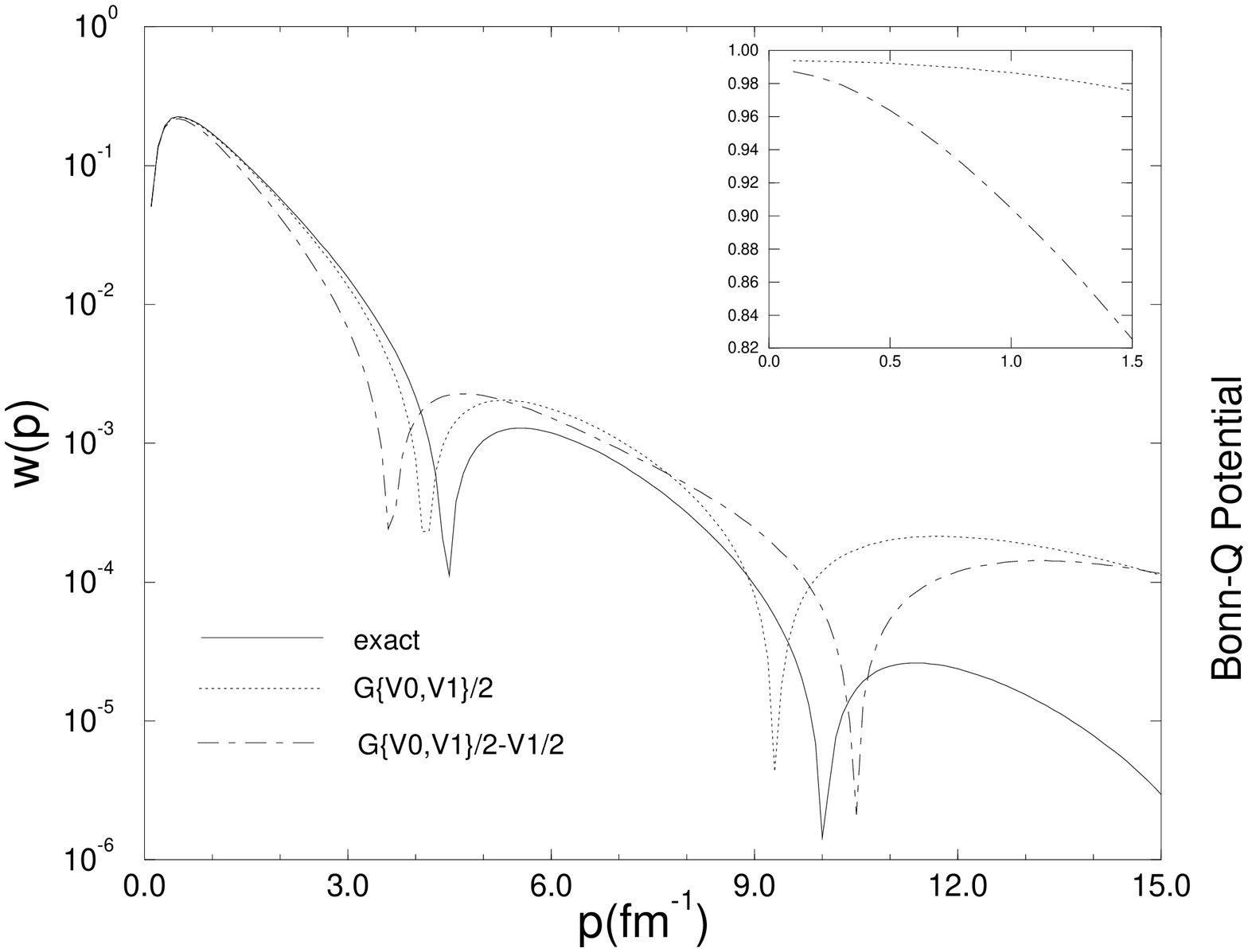, angle=270, width=10cm} }
\mbox{     \epsfig{file=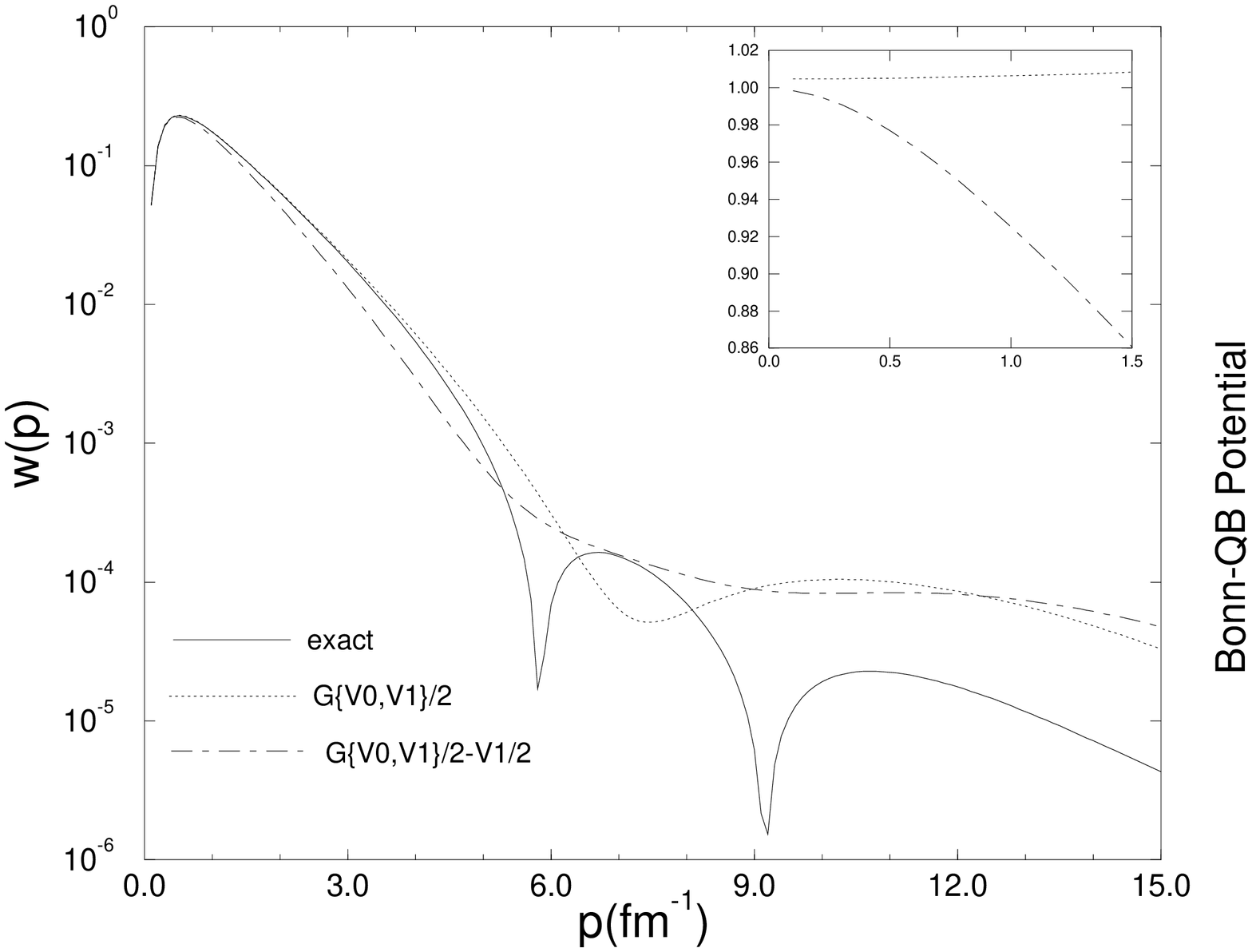, angle=270, width=10cm}  }
\vspace{-0.5cm}
\end{center}
\caption{\small{Same as in Fig. 4, but for the D wave.}}
\end{figure}  

In Figs. 4 and 5, we show the results of a perturbative calculation tending 
to estimate the effect of corrections to the potential due to the 
extra term, $\frac{1}{2} \{ V_0 , V_1 \}$ (dotted line). In this order, we 
insert as inputs in Eq. (21) the wave functions and potentials corresponding 
to the Bonn-Q and Bonn-QB NN interaction models. The resulting wave 
functions for these two models and for both S and D waves are drawn in 
Figs. 4 and 5, together with the unperturbed ones. It is 
observed that the global structure of the wave functions remains 
essentially unchanged for the Bonn-Q model and the S wave of the 
Bonn-QB model. The position of the minima 
is only slightly shifted. Quantitatively, the most important feature 
concerns the decrease of the S wave at low momenta. This one is largely 
due to a decrease of the contribution produced by the tensor force 
acting on the deuteron D wave. The effect, which is better seen in 
the insets of Figs. 4 and 5, is given by  a factor 0.73 for the Bonn-Q model 
and 0.63 for the Bonn-QB model. It shows that the wave function accounting 
for the renormalization of the interaction by the implicit 
meson in flight probability, see Eq. (13), strongly differs from the original 
one in a momentum domain, $p \simeq \kappa \; (=0.232 \, {\rm fm}^{-1})$, which 
is essential for the deuteron description. In all presented calculations, 
the D wave at small momenta is much less affected, resulting in an 
enhancement of this one when the complete perturbed wave function, 
Eq. (21), is normalized to unity. The comparison of the results 
obtained with the two Bonn models does not show significant qualitative 
differences at low momenta, despite the fact that the model Bonn-Q does badly 
for observables in relation with the tensor force. This rough agreement 
indicates the degree of reliability one can give to the  
estimate of the effects under consideration in the present work.

\begin{figure}[htb!]
\begin{center}
\mbox{\epsfig{file=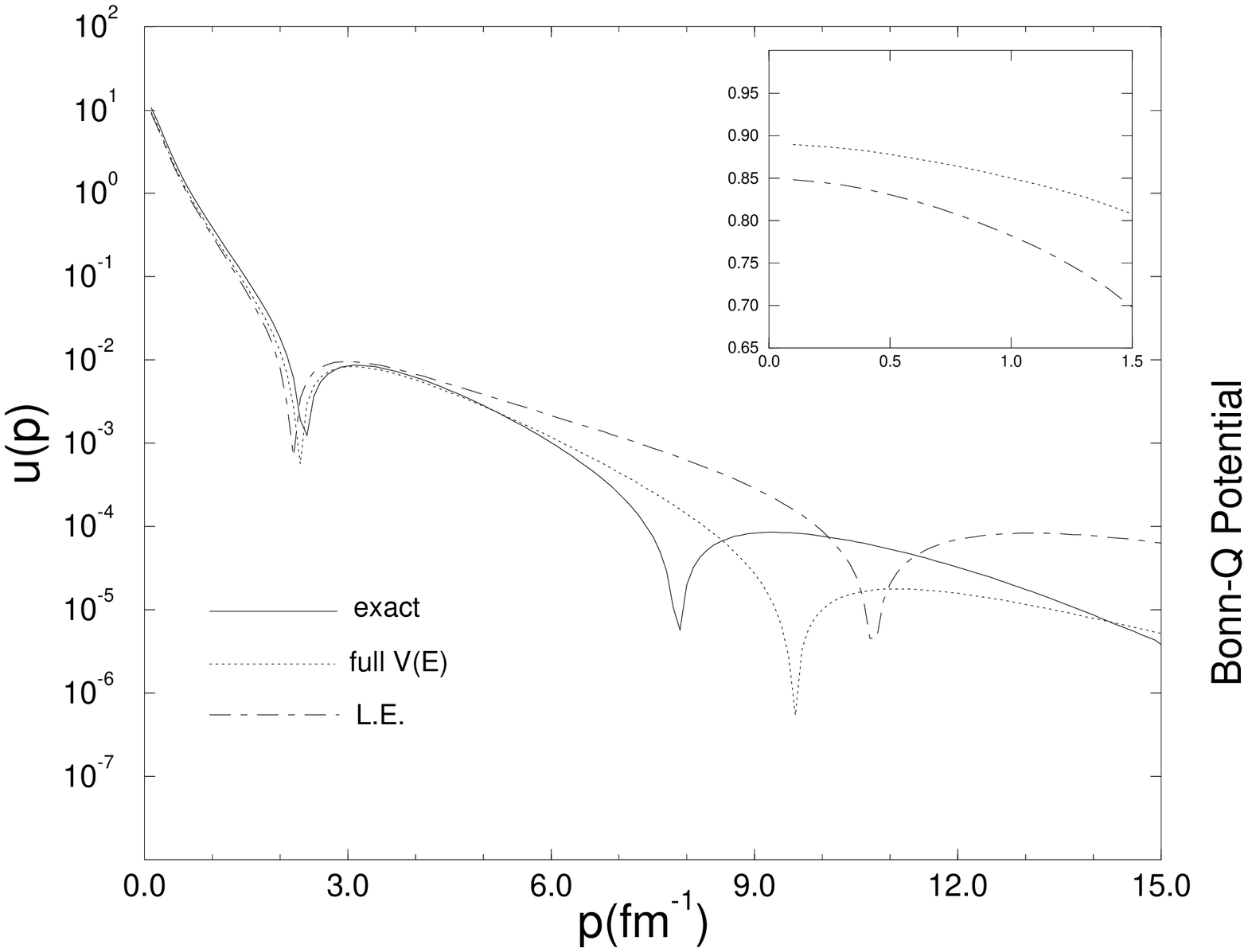, angle=270, width=10cm}  }
\mbox{\epsfig{file=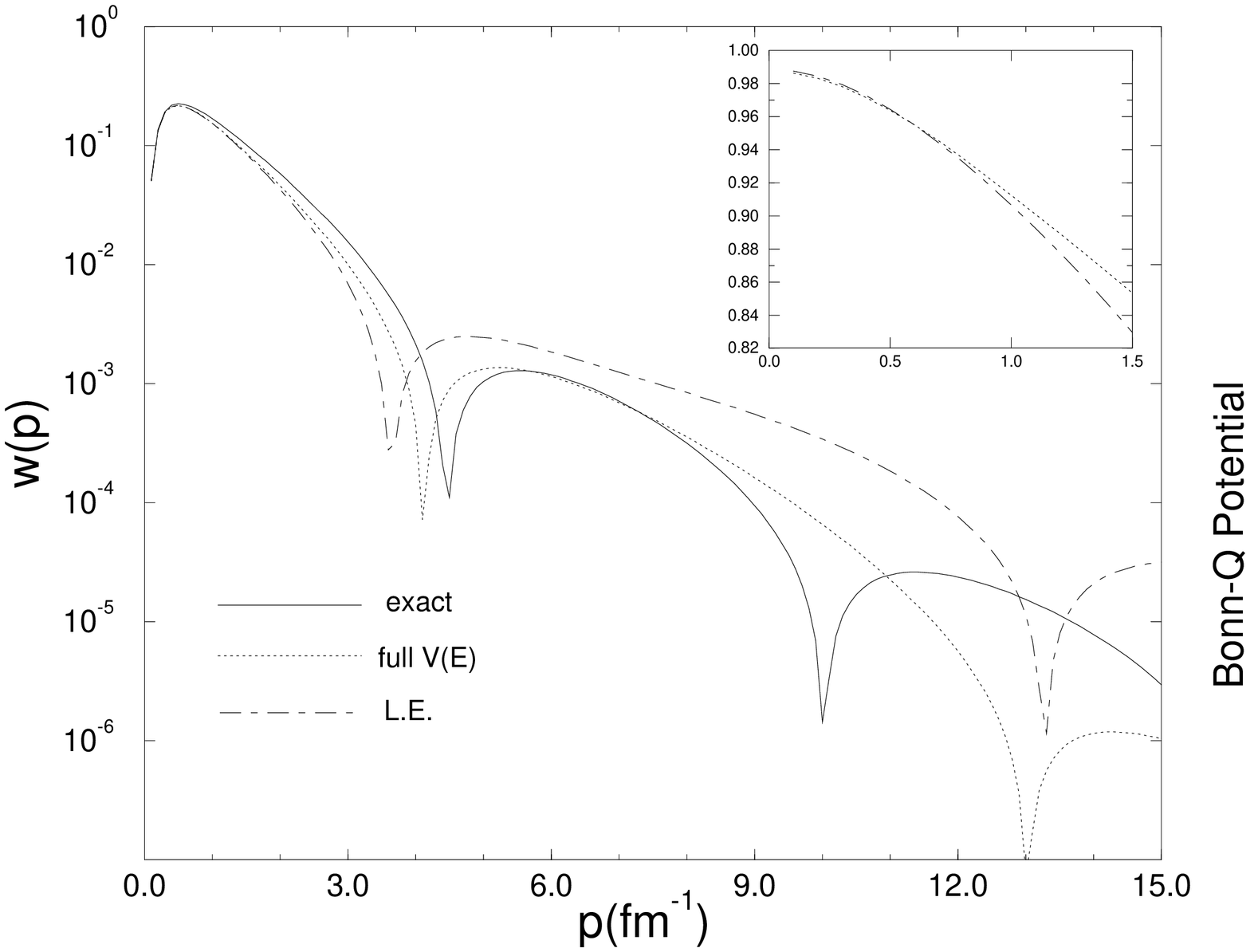, angle=270, width=10cm} }
\vspace{-0.5cm}
\end{center}
\caption{\small{Effect of the  energy dependent part of the potential, Eq. (1), 
on the calculation of the S and D components of the deuteron wave 
function  with various approximations: full kernel (dotted line, $V_E$) 
and limited expansion up to the first order in $V_1$ 
(dash-dotted line, L.E.). The input wave function (continuous line) 
and the potential are those of the Bonn-Q model. The S and D components 
are presented respectively in the upper and lower parts of the figure. 
The low momentum part is shown in the insets as a ratio to the exact results.}}
\end{figure}

The comparison of present results with those of refs. \cite{CARB2,FRED}
evidences some difference: their suppression of the S wave, roughly 
a factor 0.8, is not as large as here. Furthermore, the ``relativistic'' 
wave functions in the first reference, $f_1$ and $f_2$, 
significantly differ from their non 
relativistic limits respectively given by the usual S and D waves: a 
new zero appears in  $f_1$  while none is seen in $f_2$ in the 
range extending to $p=1.5$ GeV/c ($\simeq 7.5 \, {\rm fm}^{-1}$). 
To identify the origin of these 
differences, which may have many sources and among them genuine 
relativistic effects not accounted for here, we did a calculation 
whose spirit is closer to theirs. Using Eq. (22), we took the energy 
dependent potential given by Eq. (1) together with the unperturbed wave 
function given by the Bonn-Q model. Calculations corresponding 
to an expansion of $V_{E}$ up to the first order term in $V_1$ 
have also been performed. They may allow one to test the validity 
of this approximation, which was made in deriving the potential 
appropriate to an energy independent picture, Eq. (13). Results 
are presented in Fig. 6 for both the S and D waves, together 
with the unperturbed wave functions. As for calculations presented 
in Figs. 4 and 5 for the same Bonn model, we do not observe major 
changes in the global structure 
of the wave function. The position of the zeros is only shifted 
a little more, as is the position of the same zeros in Fig. 3 for 
the Bonn-QB model. Quantitatively, the S wave at low momenta is also 
reduced, by  a factor 0.89 for the full operator $V_{E}$ and a 
factor 0.85 for its expanded form limited to the first order 
term in $V_1$. The decrease compares to that obtained in 
refs. \cite{CARB2,FRED}, but is slightly smaller. We checked 
that this can be ascribed for a large part to the kernel of 
the interaction in the underlying light-front approach, which 
has some similarity with ours, Eq. (1), but also some difference. 
In any case, the decrease is significantly less than the one 
obtained previously in the energy dependent scheme (0.73), 
suggesting a bias in one of the approaches whose results are compared.

The difference in the reduction can be traced back to the extra 
contribution due to the term $\frac{1}{2}\,V_0\,V_1$,  which is 
present in Eq. (21), but absent in Eq. (24). This term is a 
priori different from the other one, $\frac{1}{2}\,V_1\,V_0$, 
which is present in both equations. They become equal to each 
other in the approximation where $V_0$ and $V_1$ are local and commute. 
This can explain the rough factor 2 difference between corrections 
calculated from Eqs. (21) and (24). As explained in the 
introduction to this chapter, it would be more appropriate 
in a calculation of the wave function for the energy dependent 
scheme to use the function, $\psi_0 \simeq (1+V_1)^{-\frac{1}{2}}\phi_0 $, 
as an unperturbed one. When this is done, 
the above discrepancy disappears, the remaining difference being 
provided by the operator, $(1+V_1)^{-\frac{1}{2}} \simeq 1-\frac{1}{2} V_1$.
Results calculated in this way at the first order in $V_1$ are 
presented in Figs. 4 and 5 (dash-dotted line). As expected, the wave 
functions calculated perturbatively in both schemes coincide at small momenta. 
The differences at higher momenta are similar to those already 
mentioned in relation with the discussion of results presented in Fig. 3
(effects of the factor  $(1+V_1)^{-\frac{1}{2}} \simeq 1-\frac{1}{2} V_1$).

\begin{figure}[htb!]
\begin{center}
\mbox{\epsfig{file=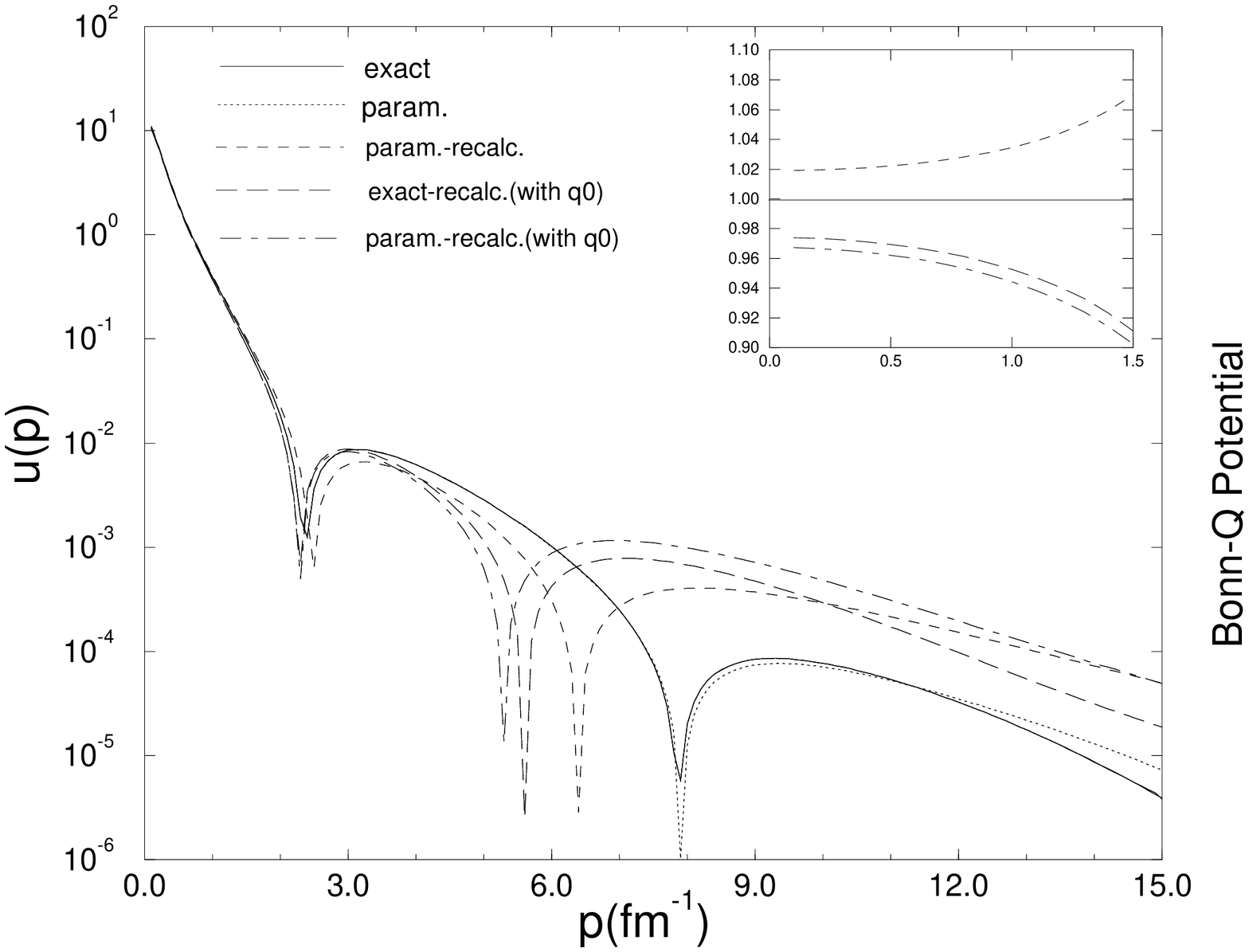, angle=270, width=10cm} }
\mbox{\epsfig{file=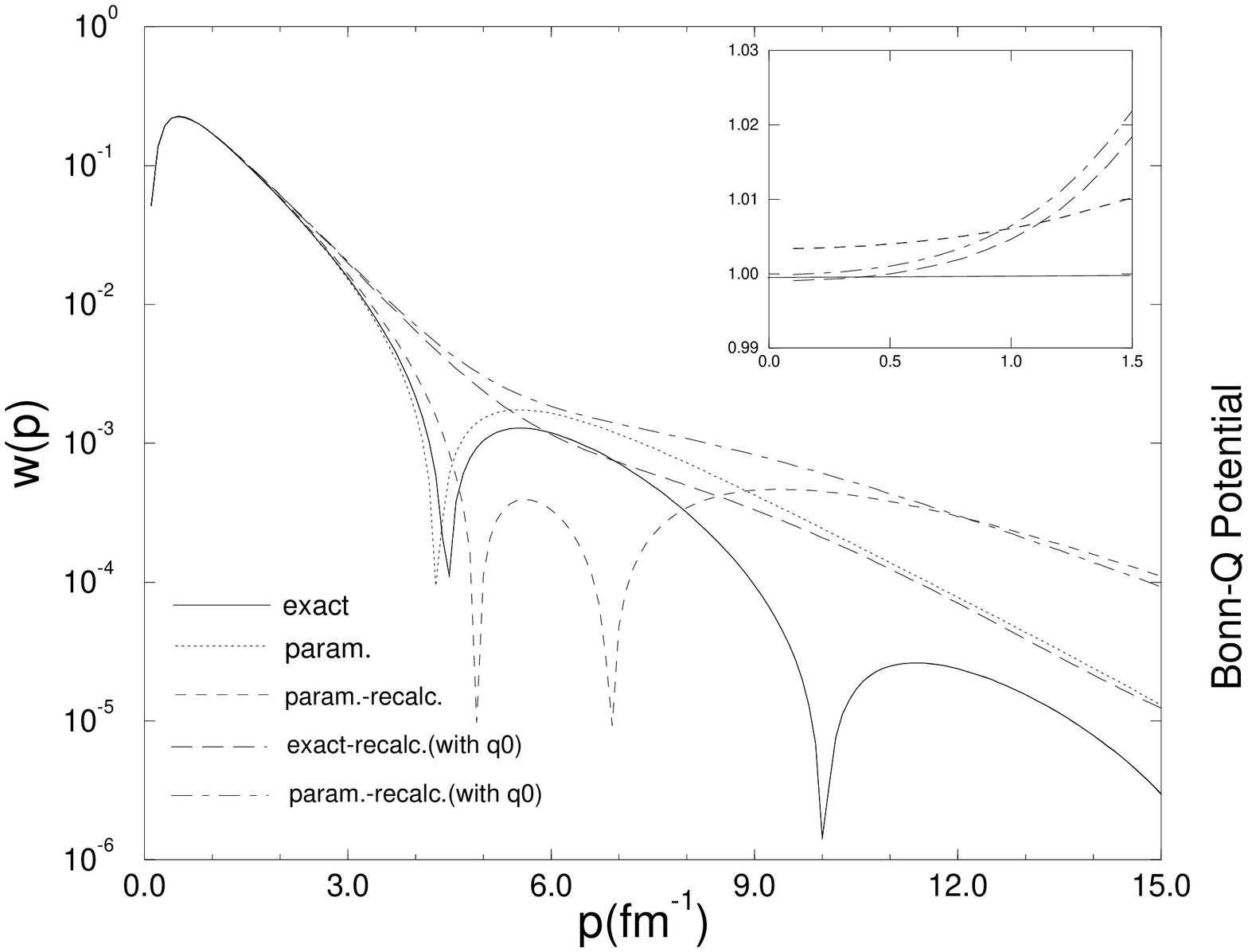, angle=270, width=10cm} }
\vspace{-0.5cm}
\end{center}
\caption{\small{Consistency check and sensitivity to the deuteron input wave 
function for the Bonn-Q model (upper and lower parts for the S and D waves 
respectively). The continuous and dotted lines respectively represent 
the exact (numerical) and parametrized wave functions (denoted exact and param. 
in the figure). The  wave function recalculated using the exact wave function 
cannot be distinguished from the original one while that one using the 
parametrized wave function (param.-recalc.), represented 
by the short-dashed line, evidences some discrepancy. The same wave functions 
recalculated using the $\rho$ meson nucleon tensor coupling employed in 
ref. \cite{CARB2} are respectively represented by the long-dash and dash-dotted 
lines (denoted exact recalc.(with q0) and param.-recalc.(with q0)). }}
\end{figure}

None of our results for the Bonn-Q model presents a qualitative 
change of the wave function similar to that obtained 
in ref. \cite{CARB2}, namely an appearance and a disappearance of 
a zero  in the S and D waves respectively (below $p=1.5 \,{\rm GeV/c} 
\simeq 7.5 \, {\rm fm}^{-1}$). As mentioned by the 
authors, this feature has been considered as a genuine feature of the 
relativistic wave functions in their approach. Surprised by the fact that 
we could reproduce other features of their calculation, especially 
the decrease of the S wave at low momenta, translating into a relative 
increase of the D wave after normalization, we performed further checks. We 
found that the parametrized wave function given in ref. \cite{BONN}, whose 
behavior at high 
momenta differs from what is theoretically expected, was not accurate 
enough to allow the authors to make a so strong statement. The 
wave functions that come out from a consistency check, Eq. (20), differ 
from the input by an amount which, in the range $p \geq 0.8$ GeV/c 
($\simeq 4.0 \, {\rm fm}^{-1}$), 
is of the same order as the effect they were interested in. 
Furthermore, the potential effectively used differs from 
the original Bonn-Q model in the treatment of the $\rho$ meson nucleon
tensor coupling. This involves a term containing the time component of the 
4-momentum carried by the meson, $q_0$. We re-calculated what should be the 
reference wave function for their study, using the parametrized wave function 
on the one hand  and their $\rho$ meson nucleon coupling. Results, which 
are presented in Fig. 7 (dash-dotted line), are qualitatively quite close to 
their so called relativistic components, $f_1$ and $f_2$. The extra zero 
in the former and the absence of a zero in the latter, in the 
range $p \leq 1.5$ GeV/c ($\simeq 7.5 \, {\rm fm}^{-1}$), thus appear as the 
consequence of approximations in 
the calculations. Other details relative to the role of various ingredients 
are shown in the figure.

\subsection{Wick-Cutkosky model, QED}
The exchange of zero mass bosons, which applies to the Wick-Cutkosky 
model or to QED, especially the Coulomb potential, deserves some 
attention. Indeed, the quantity, $V_1(r)$, defined by Eq. (9) is 
then scale independent and, being proportional 
to $\int dx ({\rm sin}\,x)\; /x^2$, is logarithmically divergent, 
requiring to come back on the definition of  $V_1(r)$.

Two questions are of interest here: the presence 
of $\alpha \, {\rm log} \, \frac{1}{\alpha}$ corrections to the binding 
energy \cite{FELD} and some correction of the order $p/m$ to 
the wave function \cite{KARM} which, both, have been attributed 
to a relativistic approach.

To determine  $V_1(r)$, we started from an $1/\omega_k$ expansion of the 
propagator $ (  E -\omega_{k} -\frac{\vec{p}_i^2}{2m} -\frac{\vec{p}_f^2}{2m}  
)^{-1}$, Eq. (3), with the idea to only retain the first order term in 
$(1/\omega_k)^2$. The presence of a singularity in the higher order 
terms when  the boson mass is zero, which implies that $ \omega_{k} 
\rightarrow 0$ for $k\rightarrow 0 $, supposes to proceed differently. 
The difficulty can be overcome by removing from the full propagator, 
Eq. (3), that part which gives rise to the usual zero mass single 
boson exchange contribution to the two-body interaction: 
\begin{equation}
\frac{1}{  E -\omega_{k} -\frac{\vec{p}_i^2}{2m} -\frac{\vec{p}_f^2}{2m}  }=
-\frac{1}{\omega_{k}}
-\frac{ E-\frac{\vec{p}_i^2}{2m}-\frac{\vec{p}_f^2}{2m} }{ \omega_k (\omega_k -E 
+ 
\frac{\vec{p}_i^2}{2m}+\frac{\vec{p}_f^2}{2m})}.
\end{equation}
Now, the extra quantity at the denominator of the second term in 
the r.h.s. of the above equation, $-E + 
\frac{\vec{p}_i^2}{2m}+\frac{\vec{p}_f^2}{2m}$, 
can be bounded from below by the absolute value of the binding energy 
of the system under consideration, $\frac{\kappa^2}{m}$, where 
$\kappa= \sqrt{m|E_{b.e.}|}$. This, which corresponds to summing 
up an infinite set of terms in Eq. (3), makes the second term in 
the above equation less singular than in the former, allowing one 
to deal with it rather safely. Being positive, the extra terms, 
$\frac{\vec{p}^2}{2m}$, don't introduce poles and, on the other hand, 
their size is expected to be the same as that of the term 
$\frac{\kappa^2}{m}$, which we will account for by multiplying 
this one by a factor 2. After making these approximations, 
the developments from Eq. (3) to Eq. (9) can be performed again with 
the result: 
\begin{eqnarray}
V_{1}(r) & = & \frac{2}{\pi}\, \alpha \, 
  \int \frac{dk k^{2}j_{0}(kr)}{\omega_{k}^2 
(\omega_{k}+2\frac{\kappa^2}{m})},\nonumber \\ 
    & = & \frac{2}{\pi}\, \alpha \, 
  \int \frac{dk j_{0}(kr)}{( k+2\frac{\kappa^2}{m})},
\end{eqnarray}
which is seen to be a function of the dimensionless variable 
$2\frac{\kappa^2}{m}r$. The coupling has been particularized to 
the case of a spinless particle (the time component for the photon) 
and the coupling $g^2$ has been expressed in terms of the 
coupling, $\alpha$, often used in QED, $g^2=4\pi\alpha$. In 
terms of this coupling, the quantity $\kappa$ for the ground 
state of the system of two equal mass particles is given by $\kappa = 
\frac{m \, \alpha}{2}$.

The small distance behavior of $V_{1}(r)$ is dominated by the 
logarithmic divergence of the integral in Eq. (27):
\begin{equation}
V_{1}(r)_{r \rightarrow 0} = \frac{2}{\pi}\, \alpha \, \left( 
1-\gamma+{\rm log}(\frac{m}{2 \kappa^2 r}) \right) +O(\frac{\kappa^2 r}{m}).
\end{equation}
where $\gamma$ represents the Euler constant, $\gamma=0.577..$. 
At very large distances ($r>>\frac{m}{\kappa^2}$), $V_{1}(r)$ 
behaves like:
\begin{equation}
V_{1}(r)_{r \rightarrow \infty} =  \alpha \, \frac{m}{2\kappa^2\,r}.
\end{equation}
A useful and approximate interpolating expression may be given by:
\begin{equation}
V_{1}(r) \simeq  \frac{2}{\pi}\, \alpha \, 
log \left( 1+ \frac{m}{2\kappa^2\,r} e^{1-\gamma} \right).
\end{equation}
For small $r$, it is exact and for $r \rightarrow \infty$, it 
provides a slight underestimate by a factor $e^{1-\gamma}/ (\pi/2)=0.97..$.

Solving the Schr\"odinger equation with the potential $V(r)$ 
together with $V_{1}(r)$ given by Eq. (28) may be done approximately 
by noticing that $V_{1}(r)$ is a smoothly varying function in the 
range $r \simeq \frac{1}{\kappa}>>\frac{1}{m}$, corresponding 
to $\alpha<<1$. It can then be approximated by a constant number:
\begin{equation}
V_{1}(r)_{r \simeq \kappa^{-1}} =   \frac{2}{\pi}\, \alpha \,  
{\rm log} \, \frac{1}{\alpha} .
\end{equation}
With the above approximation, the correction to the potential given by   
$\frac{1}{2}\{V_{1}(r),V_{0}(r)\}$ can be incorporated in the Coulomb potential, 
$V_0(r)=-\frac{\alpha}{r}$, whose coupling is renormalized as follows:
\begin{equation}
\alpha \rightarrow \alpha \left(1 -\frac{2}{\pi}\, \alpha \,  
{\rm log}\,\frac{1}{\alpha} \right).
\end{equation}
The correction to the binding energy at the lowest order immediately follows: 
\begin{equation}
E_{b.e.}= -m\, \frac{\alpha^2}{2n^2}\; \rightarrow E_{b.e.}= 
-m \, \frac{\alpha^2}{2n^2} \, \left(1-\frac{4}{\pi} \alpha 
\, {\rm log} \, \frac{1}{\alpha} \right).
\end{equation}

In all the previous discussion, we have been mainly concerned with the 
largest correction, of the order $\alpha \, {\rm log} \, \frac{1}{\alpha}$. 
We have not paid much attention to the precise value of some 
factors. Quantities such as $\frac{\vec{p}^2}{m}$ or $r$ have been 
replaced by approximate values in getting Eqs. (31, 32), which may 
affect the coefficient of $\frac{1}{\alpha}$ in the log term. 
This can be seen to correspond to a correction to the binding 
energy of the relative order $\alpha$ in Eq. (33), but does not 
change the correction of the order $\alpha \, {\rm log} \, \frac{1}{\alpha}$, 
which dominates in the limit $\alpha \rightarrow 0$. Going beyond 
requires some care, but one can anticipate that the replacement  
of the factor $2\frac{\kappa^2}{m}$ appearing  in Eqs. (27-30) 
by the potential $\alpha/r$, would provide a useful starting point for 
a discussion. With this approximation, motivated by the origin of the factor, 
$-E+\frac{\vec{p}^2}{m} $, which, acting on the wave function, provides the 
potential $\alpha/r$, $V_{1}(r)$ becomes a constant. 
Expressions with increasing accuracy would be successively given by:
\begin{eqnarray}
V_{1}(r) & \simeq & \frac{2}{\pi}\, \alpha \,{\rm log} \, \frac{1}{\alpha} \;\;
[{\rm up} \; {\rm to} \; 20\% \;{\rm for}\; \alpha \leq 0.1], \nonumber \\
& \simeq & \frac{2}{\pi}\, \alpha \, (1-\gamma+ {\rm log} \, \frac{1}{\alpha}) 
\;\;
[{\rm up} \; {\rm to} \; 25\% \; {\rm for} \; \alpha \leq 0.5], \nonumber\\
& \simeq &\frac{2}{\pi}\, \alpha \, 
{\rm log} \left( 1+ \frac{e^{1-\gamma}}{\alpha}  \right) \;\;
[{\rm up} \; {\rm to} \; 4\% \; \forall  \; \alpha ]  .
\end{eqnarray}
The first expression is identical to Eq. (31), which was obtained 
on a slightly different basis. The second and third ones could be 
used to improve Eqs. (32) and (33), when the first one is out of its 
range of validity (limited to $ \alpha \leq 0.1$). A further improvement 
consists in expressing the quantity, $2\frac{\kappa^2}{m}$, in terms of the 
renormalized potential, $ \alpha_{eff} /r$. The effective coupling, $ 
\alpha_{eff} $, which should replace the factor $ \alpha $ in the log term 
appearing in Eq. (34), is determined by the equation:
\begin{equation}
\alpha_{eff} =  \frac{\alpha}{ 1 +  \alpha \, J}
 =  \alpha (1 - \alpha_{eff} \, J),
\end{equation}
where $J=\frac{2}{\pi}\, {\rm log} \, \frac{1}{\alpha_{eff}} $, 
       $\frac{2}{\pi} \, ( 1 -\gamma +{\rm log} \, \frac{1}{\alpha_{eff}}) $, 
$\frac{2}{\pi} \, {\rm log} ( 1+ \frac{e^{1-\gamma}}{\alpha_{eff}} ) $ or a 
better expression depending on the desired accuracy.
 
The correction to the binding energy in Eq. (33) is identical 
to that found in \cite{FELD}. This result, which has been 
obtained here in a picture that is not especially a relativistic 
one, tends to show that the correction has essentially its source 
in the field theory which underlies the present approach as 
well as that followed by Feldman et al. \cite{FELD} and many 
other authors \cite{KARM,FRED}. Results for the binding energy 
of the lowest states calculated with the renormalized interaction 
have also been obtained \cite{THEU}. They explain the bulk of the 
departures from the instantaneous approximation results derived 
either from the Bethe-Salpeter equation \cite{SILV} 
or from energy dependent approaches \cite{BILA,JI,CARB1}.

It is also interesting to consider various wave functions corresponding to 
diffe\-rent approximations or different schemes. In comparison with the NN 
interaction case discussed previously, one can go a step further here by using 
the renormalized interaction which, in a first approximation, differs from the 
bare one by changing the coupling $ \alpha $ into $ \alpha_{eff} $, Eq. (35). 
In Fig. 8, we present wave functions pertinent to the lowest 
s-state. These are the usual non-relativistic wave function:
\begin{equation}
\phi_0(\vec{p})= \sqrt{4 \pi}
\frac{4 \, \kappa_0^{\frac{5}{2}} }{(\kappa_0^2+ p^2)^2} \;\;
{\rm with} \; \kappa_0=\frac{1}{2} m \, \alpha ,  
\end{equation} 
the non-relativistic wave function with the renormalized interaction:
\begin{equation}
\phi(\vec{p})= \sqrt{4 \pi}
\frac{4 \, \kappa^{\frac{5}{2}} }{(\kappa^2+ p^2)^2} \;\;
{\rm with} \; \kappa=\frac{1}{2} m \, \alpha_{eff} ,  
\end{equation} 
the wave function corresponding to the energy dependent scheme obtained 
by using Eq. (11), 
$\psi_0=(1+V_1)^{-\frac{1}{2}}\phi_0$:
\begin{equation}
\psi_0(\vec{p})= \sqrt{ \frac{\alpha_{eff}}{\alpha} } \, \phi(\vec{p}),
\end{equation} 
and the same wave function calculated perturbatively using Eq. (22), 
of which  analytic expression is given by:
\begin{eqnarray}
\tilde{\psi}(\vec{p})=\sqrt{4 \pi}
\frac{4 \, \kappa^{\frac{5}{2}} }{(\kappa^2+ p^2)^2} \; 
\sqrt{ \frac{\alpha_{eff}}{\alpha} }\; \hspace*{7.5cm} \nonumber \\
\times  \frac{\alpha}{\alpha_{eff}}
\left( \frac{1 -\frac{5\kappa^2}{4m^2} - \frac{p^2}{4m^2}
                    +   \frac{2}{\pi \,p} \,  
                  ( \frac{3\kappa^2-p^2}{2m} +                                   
                       \frac{3\kappa^2+p^2}{2m} \, \frac{\kappa^2+p^2}{4m^2} ) 
                   \, {\rm arctg} (\frac{p}{\kappa}) }{
     (1-\frac{p}{m} + \frac{\kappa^2+p^2}{4m^2} ) 
     (1 + \frac{p}{m} + \frac{\kappa^2+p^2}{4m^2})} 
           \;\;\;\;\;\;\;\;\;\;\;\;\;\;\;\;\;   \right.      \nonumber \\
   \left. + \frac{\kappa}{\pi m} \;
     \frac{2 -\frac{\kappa^2+ p^2}{4m^2} }{
       (1-\frac{p}{m} + \frac{\kappa^2+p^2}{4m^2} ) 
       (1 + \frac{p}{m} + \frac{\kappa^2+p^2}{4m^2}) } \; {\rm log} \,2 
       \hspace*{2.5cm}\right.        \nonumber \\
   \left.    + \frac{\kappa}{\pi \,p} 
       \left[ \frac{ \sqrt{ 1-2\frac{p}{m} - \frac{2\kappa^2+p^2}{m^2} }
        }{ 1 + \frac{p}{m} + \frac{\kappa^2+p^2}{4m^2} }
        {\rm log}(\frac{ 1-\frac{p}{m} +\sqrt{1-2\frac{p}{m} -                    
          \frac{2\kappa^2+p^2}{m^2} }}{\sqrt{2 \, \frac{\kappa^2+p^2}{m^2}}})    
               \;\;\;\;\;\;\;\; \right. \right. \nonumber \\
   \left. \left.  - \frac{ \sqrt{ 1+2\frac{p}{m} - \frac{2\kappa^2+p^2}{m^2} }
        }{ 1 - \frac{p}{m} + \frac{\kappa^2+p^2}{4m^2} }
        {\rm log}(\frac{ 1+\frac{p}{m} +\sqrt{1+2\frac{p}{m} -                    
          \frac{2\kappa^2+p^2}{m^2} }}{\sqrt{2 \, \frac{\kappa^2+p^2}{m^2}}})
      \right] \right),
\end{eqnarray}
where the square root (and log) functions should be appropriately changed 
when their argument gets negative (or imaginary). 

In writing Eq. (39), we have emphasized at the first line factors 
which together form the unperturbed wave function, $\psi_0(\vec{p})$, 
see Eqs. (37) and (38). The factor that accounts for the improvement 
obtained by using the complete kernel of the interaction, Eq. (1), 
is given at the following lines. For the numerical study, whose 
results are presented in Fig. 8, the value of $\alpha$ has been 
taken to be 0.5, which corresponds to $\alpha_{eff}$= 0.32. 

\begin{figure}[htb!]
\begin{center}
\mbox{\epsfig{file=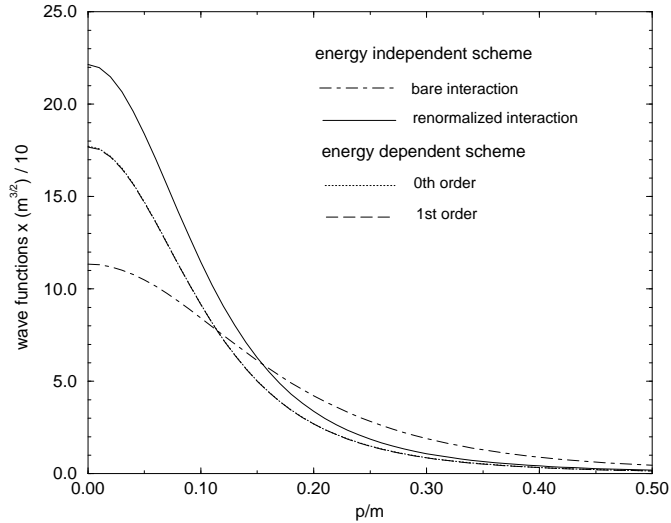, angle=270, width=10cm} }
\vspace{-0.5cm}
\end{center}
\caption{\small{Wave functions for the Wick-Cutkosky model in different 
approximations: from the bare Coulomb potential, Eq. (36) (dash-dotted line); 
from the renormalized potential, Eq. (37) (continuous line); zeroth 
order for the energy dependent scheme, Eq. (38) (dotted line); and the 
corresponding one from a first order perturbative calculation, Eq. (39) 
(dashed line, essentially on top of the previous one)). Units 
are expressed in terms of the constituent mass, $m$. The numerical 
values of the wave functions, normalized as in Eqs. (36, 37), have 
been divided by 10.}}
\end{figure}

Not surprisingly, $\phi_0(\vec{p})$ and $\phi(\vec{p})$ differ 
significantly, showing the indirect importance of the energy 
dependence of the interaction, $ V_{E}$, which was accounted for 
by employing a renormalized interaction in the energy independent 
scheme. The wave function appropriate to the energy dependent scheme, 
$\tilde{\psi}(\vec{p}) $, is quite close to the corresponding 
unperturbed one, $\psi_0(\vec{p})$. The discrepancy, which can hardly 
be seen in the figure, is less than 1\% in the domain of momenta 
covering 95\% of its contribution to the normalization. 
This indicates that the way $\alpha_{eff}$ has been derived is 
essentially correct. These wave functions strongly differ from 
that one obtained  with the bare potential in the energy independent 
scheme, $\phi_0(\vec{p})$. A similar result was obtained in the light front 
approach in refs. \cite{DESP2,CARB1} and most probably with the trial wave 
function employed in ref. \cite{JI}. The point is of relevance for 
works that have some relationship to the present one \cite{KARM}, 
where only an effect in the high momentum range was emphasized. 
This second effect, which implies corrections of 
the order $p/m$, is also present here (see below). It certainly explains 
some discrepancy between $\tilde{\psi}(\vec{p}) $ and its unperturbed 
counterpart $\psi_0(\vec{p})$, but the effect shows up at high 
momenta, outside the range shown in Fig. 8 (4\% at $\frac{p}{m}=0.5$, 
10\% at $\frac{p}{m}=0.8$ 
and 20\% at $\frac{p}{m}=1.3$). We also looked at the contribution that the 
operator $ V_1 $ provides for the normalization, see Eq. (18). The 
operator has been chosen according to calculations presented in 
this section. It is given by Eq. (27) together with the full propagator 
appearing in Eq. (26) (instead of $-\frac{\partial V_E}{\partial E}$,
which it is equal to in some limit). Discarding any theoretical improvement, 
the contribution was found to be close to what 
was approximately expected, $1-\frac{ \alpha_{eff} }{ \alpha} $, 
(0.345 instead of 0.360). This indicates, in a particular case, 
that the derivation of the effective interaction, $V(r)$, Eq. (13), its 
solution, $\phi(\vec{r})$, Eq. (12), the relation of 
this solution with that for the energy dependent scheme, $\psi(\vec{r})$, 
Eq. (11), and the normalization condition, Eq. (18) are essentially 
consistent with each other. The remaining discrepancies, which are 
at the level of a few \% in comparison with the major effect 
of 30-40\% related to the renormalization of the interaction, are 
ascribed to the approximations made in dealing with the denominator of the 
second term in the r.h.s. of Eq. (26).

We now consider more closely how the wave function of the energy dependent 
scheme given by Eqs. (1, 10) relates to the energy independent one given by 
Eqs. (12, 14) in the high momentum range. This relation in the configuration 
space,
\begin{equation}
 \psi(\vec{r})=(1+V_{1}(r))^{-1/2} \, \phi(\vec{r}) \simeq (1-\frac{1}{2} \,     
  V_{1}(r)) \, \phi(\vec{r}),
\end{equation}
reads in momentum space:
\begin{equation}
 \psi(\vec{p}) \simeq \phi(\vec{p})  -\frac{1}{2} \, 
 \int \frac{ d\vec{p\,}' }{(2\pi)^3}\; V_1(\vec{p},\vec{p\,}') \;                
 \phi(\vec{p\,}') \simeq \phi(\vec{p}) 
 +\delta \phi(\vec{p}) , 
\end{equation}
The high momentum regime of this equation is obviously that one where 
relativistic effects are expected to show up. 

Neglecting a possible renormalization of the interaction as given by 
Eq. (13), or assuming it can be accounted for effectively using Eq. (34), 
the solution of the Schr\"odinger equation for a Coulomb or Yukawa type 
interaction, 
$\phi(\vec{p})$, obeys the equation: 
\begin{equation}
\phi(\vec{p})=\frac{4\pi\alpha \, m}{p^2+\kappa^2} \int \frac{ d\vec{p\,}' 
}{(2\pi)^3} \; 
\frac{1}{(\vec{p}-\vec{p\,}')^2 + \mu^2} \; \phi(\vec{p\,}') ,
\end{equation} 
while $V_1(\vec{p},\vec{p\,}')$ is given by:
\begin{equation}
V_1(\vec{p},\vec{p\,}') = 
 \frac{4\pi \alpha}{  ((\vec{p}-\vec{p\,}')^2 + \mu^2)^{ \frac{3}{2} }  } .
\end{equation}
In the limit of large $p$, Eq. (42) indicates that $\phi(\vec{p})$ behaves like:
\begin{equation}
 \phi(\vec{p})_{p \rightarrow \infty}=\frac{4\pi\alpha \, m}{p^4} \int \frac{ 
d\vec{p\,}' }{(2\pi)^3} \; \phi(\vec{p\,}') ,
\end{equation}
where, up to some factor, the integral represents the configuration space wave 
function at the origin. The second term in Eq. (41) can be similarly calculated, 
using the expression (43) of $V_1(\vec{p},\vec{p\,}')$:
\begin{equation}
\delta\phi(\vec{p})_{p \rightarrow \infty}=-\frac{1}{2} \, \frac{4\pi\alpha 
}{p^3} \int \frac{ d\vec{p\,}' }{(2\pi)^3} \; \phi(\vec{p\,}') .
\end{equation}
Gathering results of Eqs. (44) and (45), one obtains:
\begin{equation}
\psi(\vec{p})_{p \rightarrow \infty} \simeq (1 - \frac{p}{2m}) \, 
\phi(\vec{p})_{p 
\rightarrow \infty} ,
\end{equation}
This result supposes that $V_1$ is small enough so that the first 
order expansion given in Eq. (40) is valid. It is actually limited 
to the range $ \kappa < p < m $ and neglects higher order effects in the 
coupling, $\alpha $. A result which is better with both respects, but 
limited to a zero mass boson, is given by Eq. (39) (see the second line).

The above effect may be compared to that one 
attributed to relativity in \cite{KARM}. In this work, the light 
front wave function was found to be expressed in terms of the 
non-relativistic one as: 
\begin{equation}
\psi(\vec{p})_{l.f.}= \frac{1}{1+\frac{|\vec{p} . \vec{n}|}{E_p}} \; 
\phi(\vec{p})_{n.r.} ,
\end{equation} 
where $\vec{n}$ represents the orientation of the light front. In the 
non-relativistic limit, where $\frac{p}{m}<1$, and after making an average over 
the 
orientation of $\vec{n}$, the above result reads:
\begin{equation}
\psi(\vec{p})_{l.f.} \simeq  (1 - \frac{p}{2m}) \; \phi(\vec{p})_{n.r.} ,
\end{equation}
which is identical to what was obtained above by an approach of which full 
relativity is absent. This indicates that the 
correction of the order $\frac{p}{m}$ is not cha\-rac\-teristic of a 
relativistic approach, but rather of the field theory which 
underlies Eqs. (1, 10, 11) on the one hand and (46) on the other. 
Relativity in this last equation would imply higher order terms in  
$\frac{p}{m}$, for instance in the replacement of $m$ by $E_p$ or in the 
non-isotropic dependence of $\psi(\vec{p})_{l.f.}$ with respect to the 
orientation of $\vec{n}$. In both cases, they will  manifest by effects 
of the order $\frac{p^2}{m^2}$, as usually expected. Notice that 
the above $\frac{p}{m}$ correction has for some part an elusive 
character. In the low momentum domain, it disappears from a 
correct definition of the norm, see Eq. (18), in the same way 
that renormalization effects do not show up in the calculation 
of various observables at the lowest order in $1/\omega^3$ \cite{DESP3}. 
It may nevertheless show up in the ultra relativistic regime 
where $\frac{p}{m}>>1$ \cite{KARM2} but, most probably, this is a feature 
pertinent to the energy dependent scheme.

\section{Extra contributions due to two boson exchanges}
We consider in this section the role of genuine two boson exchanges 
with respect to the correction brought about by the term in Eq. (14). 
We derive first the general expression for these contributions and 
subsequently look at particular cases depending on whether bosons are 
neutral, charged or massless. 
\subsection{General case}

\begin{figure}[htb]
\begin{center}
\mbox{ \epsfig{ file=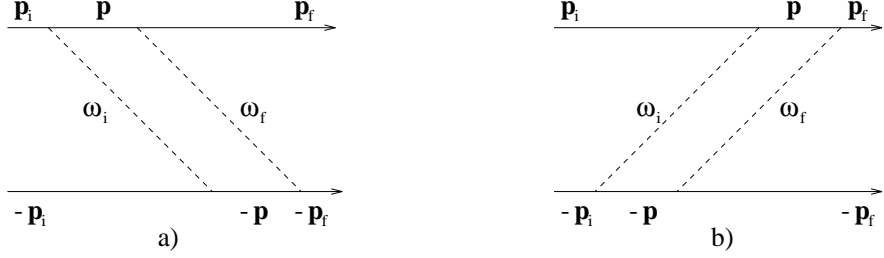, angle=270, width=13cm}}  
\end{center}
\caption{\small{Selected contributions to the two-body interaction due to two 
boson 
exchange (box-type diagrams). Apart from the common factor $  1/ (\omega_i 
\omega_f (\omega_i+\omega_f) )$, they involve a factor $1/ \omega_i \omega_f $. 
}}
\end{figure}  

\begin{figure}[htb]
\begin{center}
\mbox{ \epsfig{ file=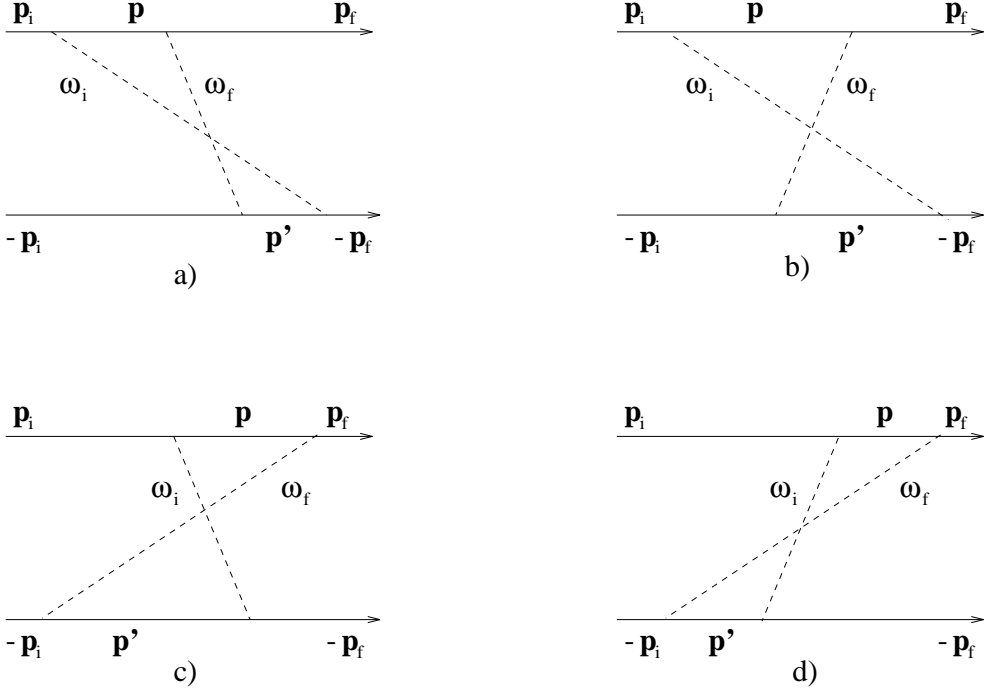, angle=270, width=13cm}}  
\end{center}
\caption{\small{Selected contributions to the two-body interaction due 
to two boson exchange (crossed-type diagrams). Apart from a common 
factor $  1/ (\omega_i \omega_f (\omega_i+\omega_f) )$, they contain 
a factor $1/ \omega_i^{2}$ (a, b) or $1/ \omega_f^{2}$ (c,d). These 
provide the dominant contribution to diagrams displayed in Fig. 2.}}
\end{figure}  

\begin{figure}[htb]
\begin{center}
\mbox{ \epsfig{ file=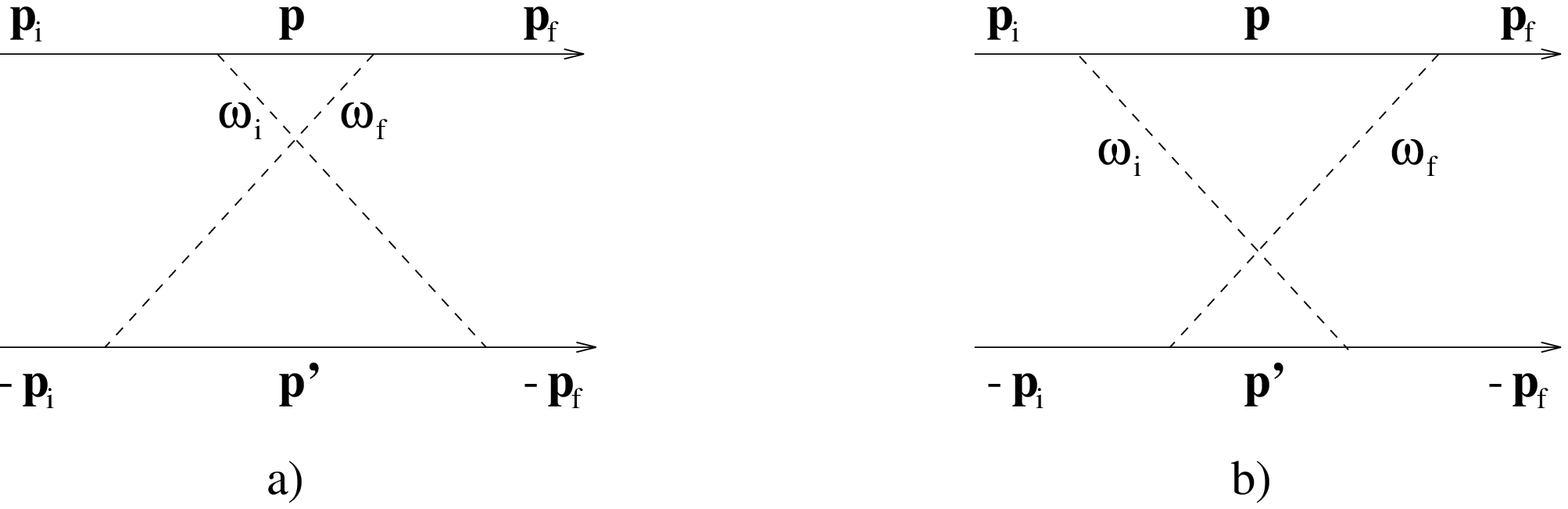, angle=270, width=13cm}}  
\end{center}
\caption{\small{Selected contributions to the two-body interaction 
due to two boson exchange (crossed-type diagrams). Apart from a common 
factor $  1/ (\omega_i \omega_f (\omega_i+\omega_f) )$, they involve 
a factor  $1/ \omega_i \omega_f $ (a,b).}}
\end{figure}

The contributions, which have the same order as the correction 
given by Eq.  (15), and are due to two boson exchanges, are 
calculated using standard perturbation theory. Only those which 
have some relevance for our purpose are retained. The corresponding 
time-ordered diagrams are shown in Figs. 9, 10, 11. They involve a 
common normalization factor, $ 1/ (\omega_i \omega_f)$ , a 
common factor corresponding to two boson propagation, 
$ 1/ (\omega_i+\omega_f) ) $, and two factors corresponding to 
single boson propagators, $1/ (\omega_i)$ or/and $ 1/(\omega_f)$.

The first ones, Fig. 9, have a box type character, and have a 
topological structure similar to that implied by Eqs. (15, 17):
\begin{equation}
\Delta V_{I}(\vec{p\,}_{i},\vec{p}_{f}) =- \frac{1}{2} \int \frac{d 
\vec{p}}{(2\pi)^3} 
\;   \sum_{v,w}  \left(  \frac{g^{2}_{w} \, g^{2}_{v} \; O^{1}_{w} \, 
O^{1}_{v} \, O^{2}_{w} \, 
O^{2}_{v}}{\omega_i^w \omega_f^v \, (\omega_i^w +\omega_f^v) } 
 \;\; \frac{1}{\omega_i^w \omega_f^v }\right). 
\end{equation}

The second ones, Fig. 10, have a crossed type character. They involve 
the two time-ordered diagrams corresponding to Fig. 1, while another 
boson is exchanged. They provide most of the contribution to the 
interaction of the two constituent particles while a boson is 
in-flight, Fig. 2. Their contribution is given by:
\begin{equation}
\Delta V_{II}(\vec{p\,}_{i},\vec{p}_{f}) =- \frac{1}{2} \int \frac{d 
\vec{p}}{(2\pi)^3} 
\;   \sum_{v,w}  \left(  \frac{g^{2}_{w} \,  g^{2}_{v} \; O^{1}_{w} \, 
O^{1}_{v} \, 
O^{2}_{v}\,O^{2}_{w}}{\omega_i^w \omega_f^v \, (\omega_i^w +\omega_f^v) } 
\;\; (\frac{1}{\omega_i^{w^2} } +\frac{1}{\omega_f^{v^2} } )\right). 
\end{equation}

The third ones, Fig. 11, have also a crossed type character. Except 
for a different spin-isospin structure, their contribution is 
identical to that shown in Fig. 9 and is given by: 
\begin{equation}
\Delta V_{III}(\vec{p\,}_{i},\vec{p}_{f}) =- \frac{1}{2} \int \frac{d 
\vec{p}}{(2\pi)^3} 
\;   \sum_{v,w}  \left(  \frac{g^{2}_{w} g^{2}_{v} \; O^{1}_{w} \, O^{1}_{v} \, 
O^{2}_{v} \, O^{2}_{w}}{\omega_i^w \omega_f^v \, (\omega_i^w +\omega_f^v) } 
\;\; \frac{1}{\omega_i^w \omega_f^v }\right). 
\end{equation}
Notice that the ordering of $O^{1}_{v}, \, O^{2}_{v}, \, O^{1}_{w}$ and 
$O^{2}_{w}$ in Eqs. (50, 51) is different from that in Eqs. (15, 49).

In deriving Eqs. (49-51), we neglected contributions involving 
the energies of the constituents, consistently with our intent 
to only consider second order effects in boson exchanges. They 
may obviously be accounted for, which supposes to extend the 
analysis we developped for one boson exchange, Eq. (1), to two 
boson exchanges.

\begin{figure}[htb]
\begin{center}
\mbox{ \epsfig{ file=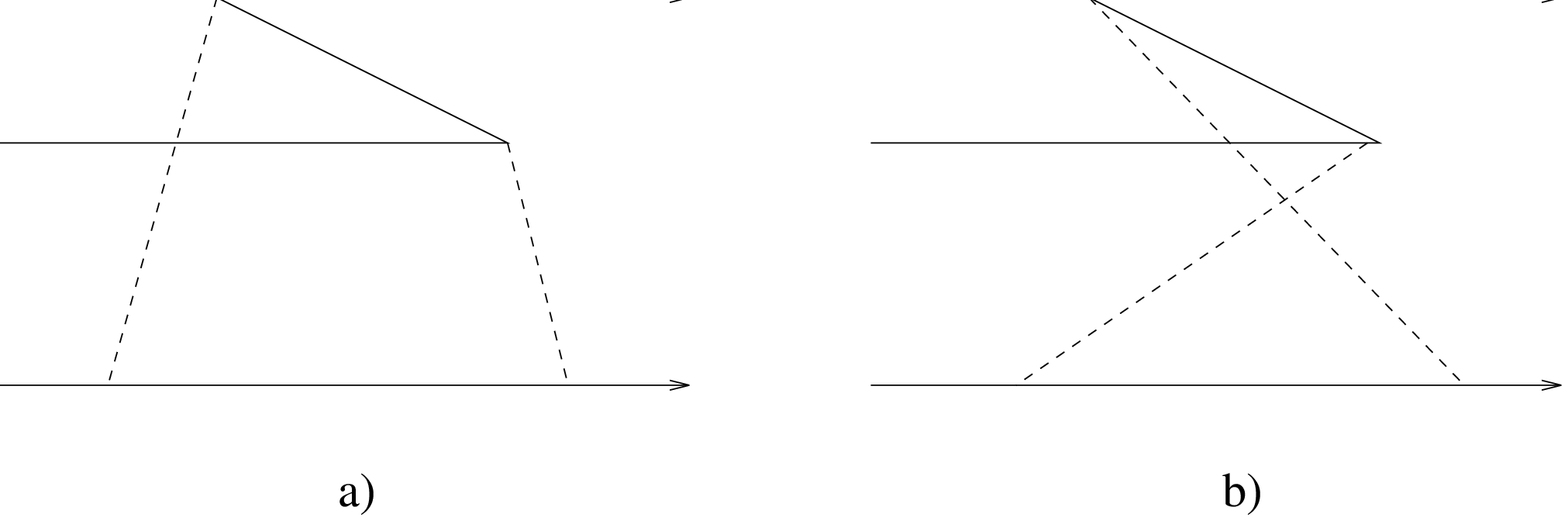, angle=270, width=13cm}}  
\end{center}
\caption{\small{Example of time-ordered two boson exchange contributions with 
negative energies for the constituents (Z-type diagrams). These 
contributions, which should be incorporated in a full description 
of the two-body interaction, don't interfer with those considered 
in this work. }}
\end{figure}

On the other hand, we did not consider the excitations of a constituent and 
its anti-particle (negative energy state). These ones, which imply Z-type 
diagrams (see Fig. 12), should be in any case accounted for, but having a 
different mathematical structure (the energy of the excited pair at the 
denominator and different vertex functions), they don't interfer with the 
discussion presented here. In the case of infinitely massive constituent 
particles and exchanged bosons coupling in a scalar way to them, their 
contribution should vanish.

The momenta in Eqs. (49-51) refer to those of particle 1, thus 
introducing an apparent asymmetry between particles 1 and 2. It 
can be verified that for the first contribution, $\Delta V_I$, the 
reference to particle 2 would be accounted for by the change of 
sign of the momenta, $\vec{p}_i, \, \vec{p}_f$ and $\vec{p}$, 
which does not provide any change of these equations. For the 
contributions produced by the crossed diagrams, $\Delta V_{II}$ 
and $\Delta V_{III}$,  Figs. 10, 11, the argument is not so 
straightforward, due to the fact that the momentum of the 
second particle in the intermediate state is not 
-$\vec{p}$, but $\vec{p}-\vec{p}_i-\vec{p}_f$ (in the c.m. system). 
The symmetry is recovered by making a change of variable.
\subsection{Neutral and spinless boson case}
The results for the case of the exchange of a neutral boson coupling 
in a scalar way and non-relativistically to constituents stem from 
Eqs. (49-51). This applies especially to the $\sigma$ and $\omega$ 
mesons coupling to nucleons and to the photon coupling to charged 
particles (Coulomb part). In both cases, their contribution provides 
a large, if not the dominant part of the interaction. Taking into 
account that the vertex functions, $O_v$ and $O_w$, commute, it is 
easy to sum up the different contributions, $\Delta V_I, \, \Delta V_{II}$ 
and $\Delta V_{III}$, with the result that the common factor, 
$(\omega_i^w +\omega_f^v)^{-1}$, cancels out:
\begin{equation}
 \Delta V_I+\Delta V_{II}+\Delta V_{III}= -\frac{1}{2} \int \frac{d 
\vec{p}}{(2\pi)^3} 
\;  \sum_{w,v} \left( \frac{g^{2}_{w} \,g^{2}_{v} \; O^{1}_{w} \,O^{2}_{w} \,  
O^{1}_{v} \,O^{2}_{v}}{\omega_i^{w^2} \omega_f^{v^2} }
(\frac{1}{\omega_i^w} +  \frac{1}{\omega_f^v} ) \right).
\end{equation} 
Remembering that, in the approximation under consideration, the 
vertex functions, $O$, reduce to unity, this total contribution 
has an attractive character. It  just cancels the contribution 
which arises from removing the energy dependence from Eqs. (1, 4, 10) 
and corresponds to the renormalized potential $V$, and the 
associated wave function, $\phi $.

This result is important in the sense that it indicates that usual 
single boson exchange potentials, that are used in the Schr\"odinger 
equation and are described by Yukawa potentials, already account, 
in a hidden way, for part of the two boson exchange contribution, 
including crossed diagrams. By the same token, they also account 
for the renormalization and recoil effects related with the energy 
dependence of the genuine interaction, $V_E$, but this may be true 
only at the lowest order in $\frac{1}{\omega^3}$.

Neglecting some non-locality, a physical interpretation of the above 
contribution to the two-body interaction may be obtained by going 
back to its expression in configuration space:
\begin{equation}
 \Delta V_I+\Delta V_{II}+\Delta V_{III}= V_1(r)\; V_0(r),
\end{equation}
where $ V_0(r)$ and $V_1(r)$ have been defined previously (Eqs. 6, 9). 
It corresponds to the interaction between the two constituents, 
$ V_0(r)$, multiplied by the boson in flight probability, which 
is given by $V_1(r)$. It is 
not a surprise therefore if it cancels the normalization correction to the 
interaction potential, $\frac{1}{2}\{V_{1}(r),V_{0}(r)\}$, 
which was corresponding to missing this contribution. Obviously, 
as already mentioned, the cancellation holds provided that the 
two constituent particles in the intermediate state with a boson 
in flight have the same spin, isospin or angular momentum as the 
initial or final states. A little bit of caution is nevertheless in order in 
identifying Eq. (53) with the contribution of Fig. 2, which at 
first sight could be written as the interaction of the two constituent 
particles, $V_{0}(r)$, multiplied by the probability to have a boson in flight. 
The argument misses the point that, in order to get $V_{0}(r)$ in Eq. (53), 
all relative times corresponding to an emission of a boson by one constituent 
and its absorption by the other have to be considered. This cannot be 
achieved by crossed diagrams alone as the time at which the absorption 
takes place (in Fig. 10a for instance) is necessarily bounded by the time 
at which the first emitted boson is absorbed. This is the place where 
the contribution of the box diagram (Fig. 9a) is relevant as it precisely 
provides the missing time ordered contribution. In view of this, one can 
imagine that the cancellation of the norm correction to the potential 
together with that one involving the interaction of the two constituents 
while one boson is in flight (part of Fig. 2) can be generalized to an 
undetermined number of bosons in flight, within the conditions under 
which  Eq. (53) has been derived (locality of the interaction, spin- and 
isospin-less particles, no Z diagram). A schematic proof is given in the 
appendix.

\begin{figure}[htb]
\begin{center}
\mbox{ \epsfig{ file=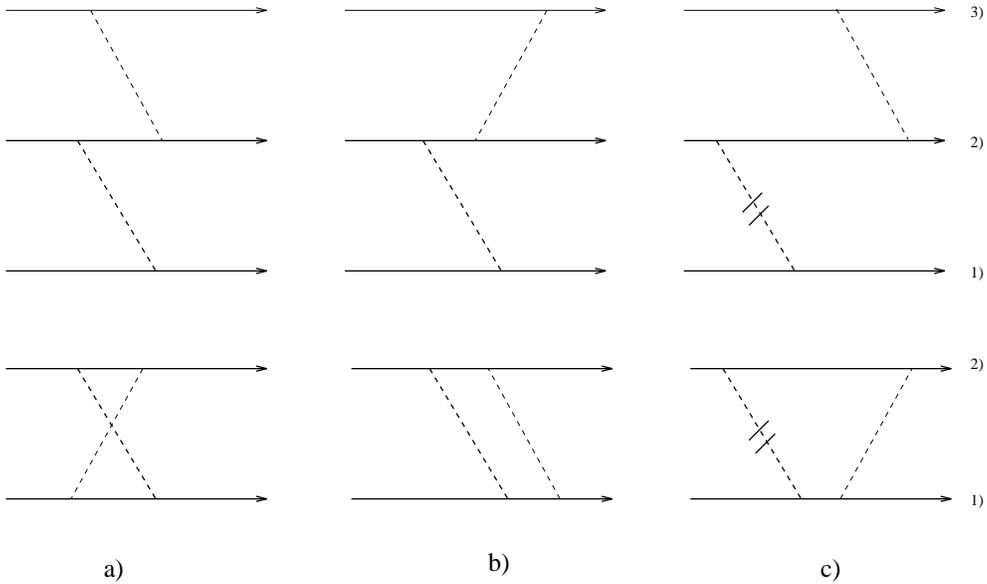, angle=270, width=13cm}}  
\end{center}
\caption{\small{Time ordered diagrams showing the relationship of two boson 
exchange 
contributions considered here (bottom part) with recoil-norm contributions 
associated to the interaction with an external probe (represented by 
a thin line in the top part). The different diagrams, a, b and c, 
correspond to crossed, box and renormalization contributions. The double 
bar on the boson line in the diagram c) reminds that only the part contributing 
to the norm has to be retained. The bottom diagrams are obtained by bringing in 
the top ones the line 3 in coincidence with the line 1.}}
\end{figure}  

The above cancellation has been obtained in the case of constituents with 
equal masses. There is no difficulty to check that it also holds with 
unequal masses within the same approximations. In the present approach, this 
is a first step in allowing one to recover in particular the usual Coulomb 
interaction of an electron (light particle) in the field of a nucleus (heavy 
particle). 

The cancellation of contributions given by Eq. (52) with those due 
to the ``renormalization'' of both the interaction and the wave 
function, Eq. (17), reminds the norm-recoil cancellation discussed 
in the literature in the 70's. This one however concerns the 
interaction of a two-body system (in the simplest case) with an 
external probe and thus involves at least three bodies. We examined 
this situation and indeed find some relationship. The difference 
with the present case is that one of the two bosons exchanged 
between the two bodies in Figs. 9, 10, 11  is attached to a third body, 
the electron for instance for an electromagnetic probe, see Fig. 13. Apart 
from a possible factor related to the number of constituents, there is 
a one to one correspondence. The demonstration is achieved by 
noticing that all time ordered diagrams are summed up while the 
two bosons attached to the same constituent line interact independently 
with the two other lines, with the consequence that the corresponding 
vertices necessarily commute.

\subsection{Massless neutral and spinless bosons}
The logarithmically divergent character of the extra contribution 
to the interaction given by Eq. (17) in the limit of a zero mass 
boson also occurs for the contribution given by Eq. (52). Like there, 
to overcome the difficulty, one has to take into account the 
binding energy in the boson propagators appearing in Eqs. (49-51). 
A priori, the corrections differ from those appearing in 
Eq. (27), preventing contributions of the two neutral boson 
exchanges to cancel that one due to the renormalization of the 
wave function. However, we saw that the domi\-nant correction 
arising from this renormalization, of relative order 
$\alpha \; {\rm log} \, \frac{1}{\alpha} $, was unaffected by the precise 
value of the coefficient to be 
inserted in front of this binding energy (2 in Eq. (27)). As 
a result, this particular correction  cancels out when considering both 
the renormalization of the wave function and the two boson 
exchange contribution.

Remaining corrections to the interaction may therefore be of relative 
order $\alpha$. At first sight, it is not clear whether these ones 
should also cancel out from the consideration of the above contributions 
alone (assuming they should cancel at all as expected from relativistic 
calculations where the correction to the dominant order is often  
believed to be of the order $\alpha^2$ rather than $\alpha$). The problem 
is that the long range of the force makes it difficult to discuss the 
contributions order by order, the binding energy introduced to remove the 
logarithmic divergence in Eq. (27) accounting for some of them. However, 
on the basis of the result presented in the appendix and assuming we 
have not missed some mathematical sublety, we expect that these 
corrections of relative order $\alpha$ with respect to the dominant 
term should vanish. It is obviously out of question to discuss here 
the next correction of relative order $\alpha^2$, 
which would also suppose to precise the framework (equation) to be used 
as well as to account for other radiative corrections.
\subsection{Residual contribution for bosons with spin or charge}
In the case of the exchange of bosons carrying some charge 
(isospin or color) or some spin, or coupling to the spin of 
constituents, the vertex functions, $O$, in Eqs. (49-51) do not 
commute. Some residual contribution then remains when 
adding the contribution of two boson exchanges, Eqs. (49-51), 
and the contribution ari\-sing from the renormalization, Eq. (17). 
It occurs for bosons currently referred to as the pion in the field of 
nuclear physics or the gluon in QCD. Such contributions are 
generally ignored. Part of them may be accounted for in a 
phenomenological approach, by fitting parameters of a potential to NN 
scattering cross sections for instance. While one cannot 
underestimate the power of this procedure, it should be 
noticed that the resulting two pion contribution has certainly 
a very complicated structure and it is not sure that it can be 
easily approximated by single meson exchanges. One can thus 
imagine that these last approaches, by trying to reproduce 
fine contributions which their simplicity does not permit 
in principle, introduce some bias in other sectors of 
the interaction. With this respect, we notice that the Paris 
model of the NN interaction is the only one where the two pion 
exchange contribution is correctly accounted for \cite{PARI}. 
It however suffers from some drawbacks, such as neglecting 
terms of the order $(p/m)^4$, which now turn out to be relevant 
in determining a parametrization of the model \cite{AMGH}. The 
two pion exchange contribution was also considered in the full 
Bonn model \cite{BONN}, but its energy dependence and the 
problems this feature are raising have led to its discarding. 
It is finally worthwhile to mention the description of the Nucleon Nucleon 
system from the Bethe-Salpeter equation \cite{TJON} or from other relativistic 
approaches \cite{GROS}. Based on single meson exchange, they incorporate 
contributions discussed in the first part of this work (Sect. 2), 
but neglect contributions corresponding to the crossed diagrams 
of Figs. 10 and 11, thus introducing another kind of bias.

We tentatively examined contributions due to the above residual two 
meson exchange contribution in the case of the NN interaction. 
The idea is to get an order of magnitude and, 
consequently, to know whether one has to worry about them. In 
this aim, we neglect the contribution of the box diagram, 
Eq. (49) and a contribution of the same magnitude for the 
crossed diagram. The terms left out are of the order 
$\omega_i \, \omega_f/(\omega_i+ \omega_f)$, where $\omega_i$ 
and $\omega_f$ refer to the on-mass shell energies of the 
two exchanged mesons. This is at most equal to 1/4 ($\omega_i =\omega_f$), 
which gives some estimate of what is neglected. Due to possible 
destructive interfe\-ren\-ces, actual corrections may be larger however. 
With the above approximation, the expression of the two meson exchange 
contributions given by Eqs. (49-51) becomes identical to that 
of the norm correction to the energy independent interaction, 
Eq. (17), except for a different ordering of the spin and isospin 
operators. Furthermore, we concentrated on the contribution of 
isospin 1 mesons, which in particular contain the pion whose 
contribution to the NN interaction is known to be important, 
and pick up the part whose spin structure is the same as for the term  
$\frac{1}{2}\{V_{1},V_{0}\}$. The residual contribution under 
consideration thus reads
\begin{eqnarray}
V_{res.}^{NN}=
- \left[ \; (\vec{\tau_1}.\vec{\tau_2}) (\vec{\tau_1}.\vec{\tau_2})
-\vec{\tau_1}\;(\vec{\tau_1}.\vec{\tau_2})\;\vec{\tau_2}\; \right]_{sym.}
\, \frac{1}{2}\{\hat{V}_{1},\hat{V}_{0}\} \nonumber \\
=4 \, (\vec{\tau_1}.\vec{\tau_2}) \, 
\frac{1}{2}\{\hat{V}_{1},\hat{V}_{0} \}, 
\end{eqnarray}
where the first and second terms in the squared bracket respectively 
arise from the renormalization of the interaction (generalization 
of Eq. (15)) and the crossed diagrams (generalization of Eq. (53)). 
The $ \hat{ } $ symbol indicates that only the isospin 1 meson 
contribution to the full interaction is considered, the corresponding 
pro\-duct of isospin matrices being factored out. 

The relative effect of the two terms is immediately obtained. For 
the deuteron, of which isospin is zero, they are respectively 
proportional to 9 and 3. This evi\-dences a constructive interference, 
contrary to the neutral spinless meson case where the interference 
was destructive and complete. Amazingly, the second term in the 
bracket on the r.h.s. of Eq. (54) involves the interaction of two 
nucleons in  $^1S_0$, $^3P_0$,.. states, which may look surprising. 
One can convince 
oneself however that nothing is wrong as this state is coupled 
to a meson in flight which, beside some angular momentum, carries 
isospin 1, allowing one to get the total expected isospin, 0.  

\begin{figure}[htb]
\mbox{ \hspace*{-6mm} \epsfig{file=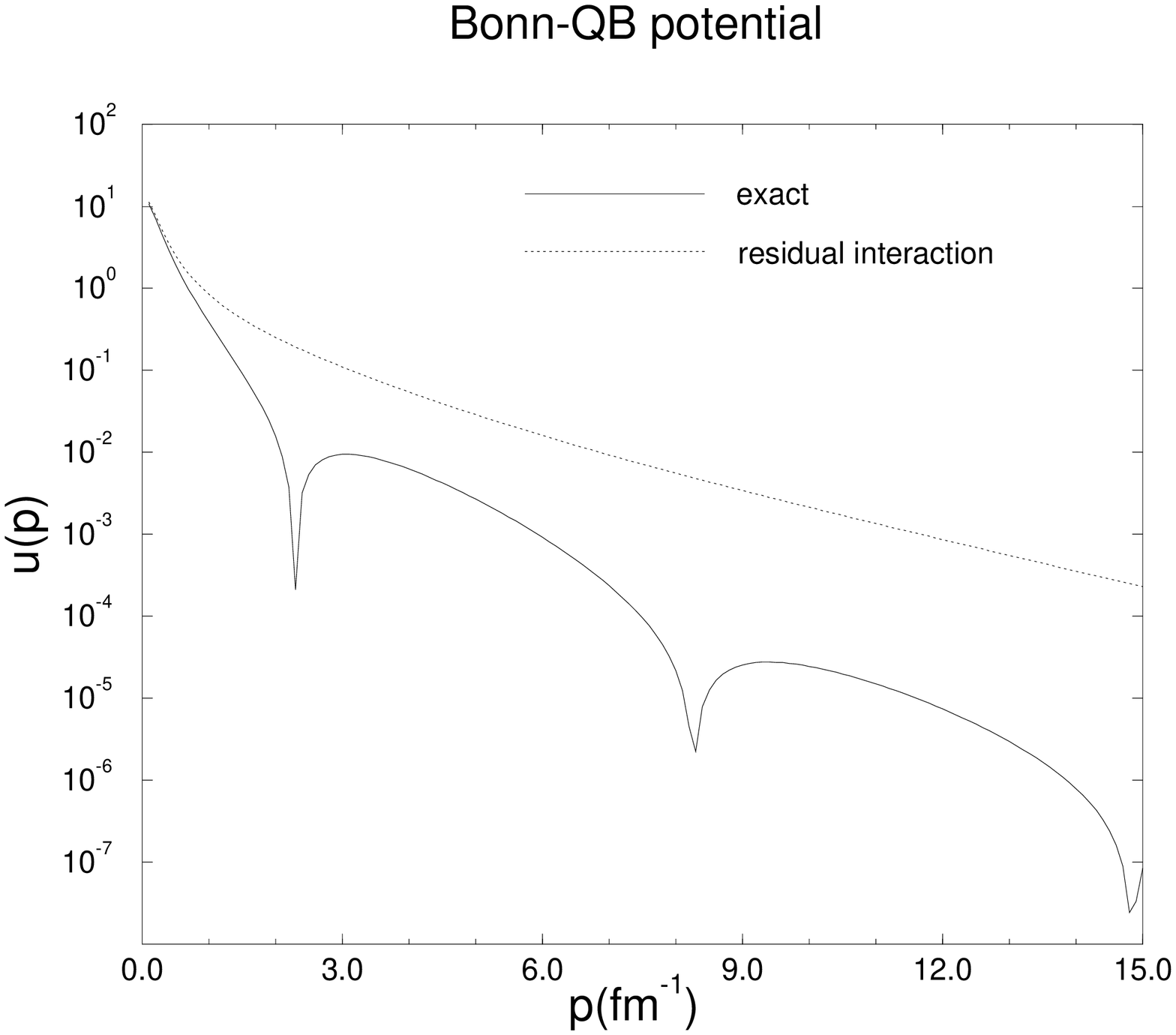, angle=270, width=7.4cm} 
 \hspace*{-9mm}      \epsfig{file=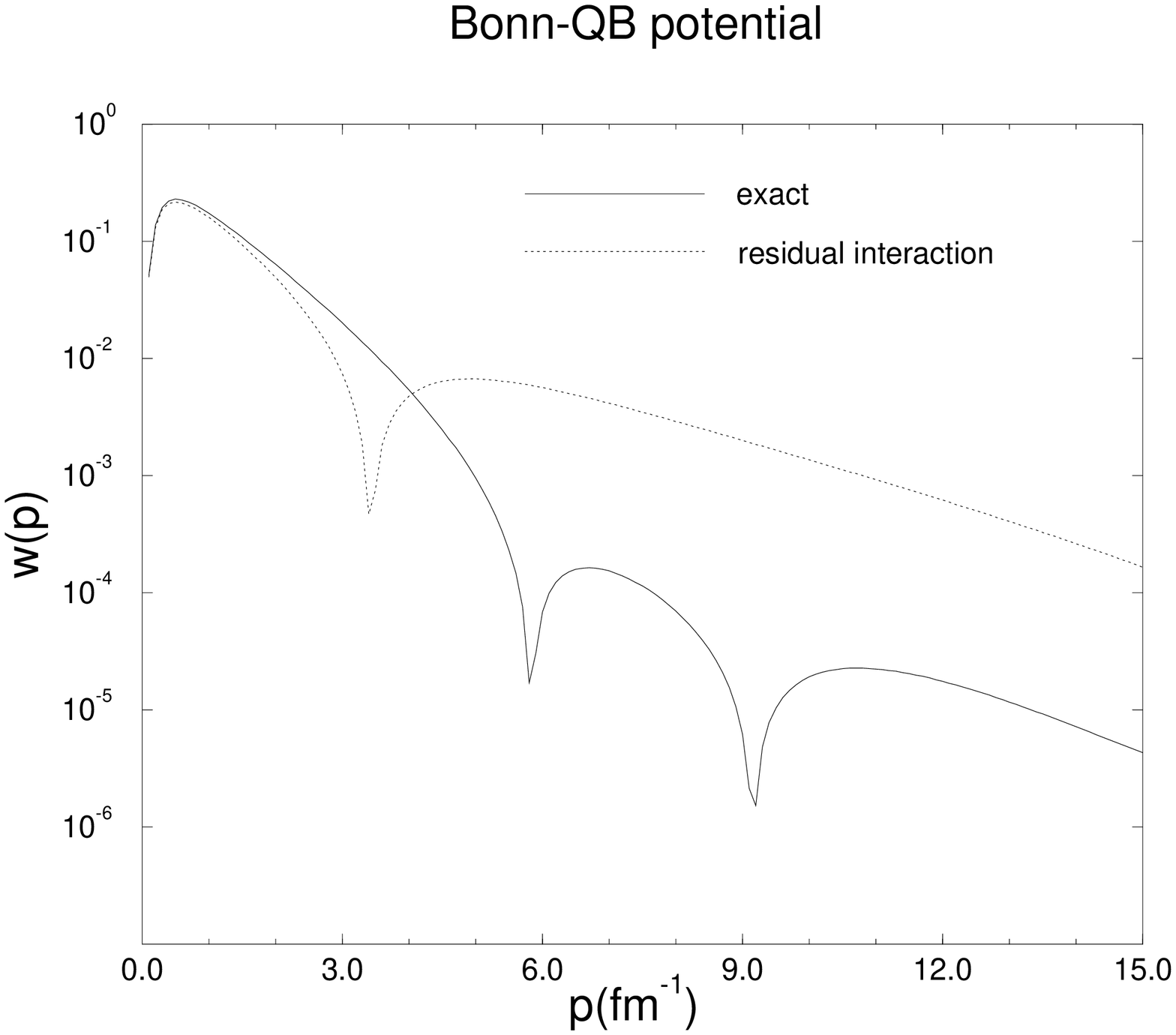, angle=270, width=7.4cm}  }
\caption{\small{Effect of a residual NN interaction due to isospin 1 mesons: 
contributions to the S~and D wave components of the deuteron state shown 
in the figure are calculated perturbatively.
}}
\end{figure}

The effect of the residual interaction, Eq. (54), can be seen in 
Fig. 14 for both the S and D wave components of the deuteron 
state (Bonn-QB model). Not much is seen at very low momentum 
for this interaction model. For the S wave, this may be accidental as a large 
effect was obtained with the Bonn-Q model, comparable to the total correction 
obtained previously from the term $\frac{1}{2}\{V_{1},V_{0}\}$ (see Figs. 4 
and 5). At intermediate momenta, the zero at 2 ${\rm fm}^{-1}$ is washed 
out in both cases while at large momenta the contribution 
blows up. It is not clear whether this is a real effect or the 
result of picking up a part of the potential in a domain where what 
comes from the exchange of mesons with different isospin is not 
well determined.

Using approximations similar to those made in getting Eq. (54), 
one can also get an expression for the residual interaction due 
to the gluon exchange (Coulomb part). It reads
\begin{eqnarray}
V_{res.}^{qq}=- \left[ \, (\vec{\lambda_1}.\vec{\lambda_2})
(\vec{\lambda_1}.\vec{\lambda_2})
-\vec{\lambda_1}\;(\vec{\lambda_1}
.\vec{\lambda_2})\;\vec{\lambda_2} \, \right]_{sym.}
\, \frac{1}{2}\{\hat{V}_{1},\hat{V}_{0}\} \nonumber \\
=6 \, (\vec{\lambda_1}.\vec{\lambda_2}) 
\, \frac{1}{2}\{\hat{V}_{1},\hat{V}_{0}\}, 
\end{eqnarray}
where the $\lambda$ matrices act in the color space. The $ \hat{ } $ 
symbol now indicates that only the Coulomb part of the interaction 
should be considered. Using Eq. (34), the above 
correction amounts to renormalize the Coulomb part of the gluon exchange by a 
factor roughly equal to 0.5 for $\alpha_s=0.5$. The size of the effect 
certainly requires a more careful and complete examination, especially in 
view of the fact that it reduces the role of the Coulomb part of the one gluon 
exchange in explaining the meson and baryon spectroscopy. It does not support 
the result of some phenomenological studies that are favoring larger 
values of $\alpha_s$, in particular in relation with the position of radial 
excitations \cite{DESP7,VALC}. 

Notice that the residual interaction, Eq. (55), involves
$\alpha \,{\rm log} \, \frac{1}{\alpha}$ corrections, thus indicating that 
they don't cancel in all cases as the example of the Coulomb interaction 
may a priori suggest. These corrections have some relationship with the well 
known log corrections that account for the renormalization of the weak 
interaction of quarks due to QCD effects.

\section{Conclusion}
In this work, we have studied the relationship of energy dependent 
and energy independent two-body interaction models. The energy dependence 
of the first ones, based on a field theory approach, has obviously some 
theoretical support but, as is well known, is also a source of 
difficulties. The energy independent models are partly motivated 
by their empirical success, starting with the success of the Coulomb 
interaction in describing the dominant part of the electromagnetic 
interaction.

Our study has concentrated on the effect of possible corrections 
to wave functions. Essentially, we showed that effects accounted 
for by an energy dependent model could be incorporated to some 
extent in an energy independent approach. The resulting interaction 
is different from the usual ones, of the Yukawa or Coulomb type. These 
ones turn out to be renormalized by the probability that the system 
under consideration be in a two-body component, excluding that part 
involving bosons in flight. Far from raising some problem, we, on 
the contrary, believe that this result provides a useful intermediate step in 
the derivation of effective two-body interaction models to be employed 
in an energy independent scheme. Indeed, beside the above contribution, 
we expect in any case some contribution due to two crossed boson 
exchanges. For neutral spinless bosons, this one cancels the 
contribution due to the renormalization of the interaction together 
with an extra smaller two boson box contribution. This result, which 
holds at the zeroth order in the inverse of the constituent masses, 
possibly different and finite, allows one to recover the well known 
Yukawa or Coulomb interactions (at least for some part in the last case). 
It confirms what has been obtained 
by other me\-thods for these interactions \cite{BREZ,TODO,GROS2,NEGH,NIEU}, 
perhaps providing a more intuitive explanation. It is not however 
clear whether the 
conditions required to obtain the cancellation are identical in all cases, 
authorizing some scepticism about the result \cite {DARE}. 
From present results, from those quoted above  \cite{NIEU} and from 
results to come concerning bound states \cite{THEU}, one can thus 
make the following conjecture for neutral spinless bosons. The description 
of low energy states obtained from solving the Bethe-Salpeter equation 
with a single boson exchange interaction as well as that obtained with 
an energy dependent interaction (including on the light front) has a 
strong relationship to a description obtained in an energy independent 
picture with a renormalized Yukawa or Coulomb interaction. Physics 
described by a Yukawa or Coulomb interaction in an energy independent 
scheme rather has a strong relationship with that obtained from the 
Bethe-Salpeter equation or an energy dependent scheme together with an 
interaction including crossed boson exchange beside single boson exchange. 
In the case of bosons with charge or spin, some 
residual contribution is left, which by no means seems to be 
negligible for hadronic systems of current interest in nuclear 
or particle physics. 

While contributions related to the energy dependence have to be 
consi\-dered and cannot be separated from these other ones allowing 
one to recover the Coulomb interaction for instance, what has to 
be done in practice with them depends on whether one looks at the 
two-body interaction from a theoretical or a phenomenological 
point of view. In the first case, they are part of a series of 
contributions that have to be determined with more or less accuracy 
depending on the domain and especially on the size of the coupling. 
In the other case, as in the NN strong interaction one, a large body 
of data  has allowed one to determine phenomenological interaction 
models, accounting for a large part of the new contributions. Dealing  
with them requires some care, beyond that one mentioned in the 
literature when employing wave functions issued from an energy 
dependent model fitted to NN scattering data 
\cite{DESP1,PAUS,AMGH1,DESP5,DESP3}. 

In the energy independent scheme and for the simplest case, this supposes 
to cancel the new contribution with another one equal to it 
but of opposite sign: nothing is changed with respect to the 
original model. If one considers that the new contribution has 
a genuine character, what we believe, then a change of the parameters 
determining the shorter range part of the models has to be performed. 
In the energy dependent models, the same can be done but in any case a 
readjustment of the parameters is required. Furthermore one has 
to keep in mind that wave functions in both schemes are not 
equivalent and cannot be used on the same footing as input as well as ouput. 
Thus, the zeroth order unperturbed wave function in an energy dependent 
scheme should not be taken as that one in an energy independent scheme as 
one may expect at first sight \cite{CARB3,FRED,CARB2}, but rather as the 
product of this one by the operator, $(1+V_1)^{-\frac{1}{2}}$,
where $V_1$ is related to the derivative of the interaction with respect 
to the energy, $ V_1=-\frac{\partial V_E}{\partial E} $.

The present work has been essentially devoted to the 
first order contributions issued from the energy dependence of 
field-theory based interaction models. In the case of a 
phenomenological approach to the NN strong interaction, they turn 
out to be largely included in these models, as already mentioned. They 
have an elusive character for some part and should not provide 
much effect in analyzing physical results at low or moderate 
energies. In view of the present work, the large effects found 
in refs \cite{CARB2,FRED} thus 
appear as resulting from accounting roughly for twice the same 
contribution. Beyond first order corrections, some effect in 
relation with boson in flight should show up. 
They are likely to require studies more refined than those presented 
here.

{\bf Acknowledgement}
One of the authors (B.D.) is very grateful to T. Frederico for discussions 
which led to the developments presented in this work and to students, P.  
Gaspard and I. Pfeiffer, whose exploratory work was quite useful for 
the present one.
We would also like to thank R. Machleidt for providing us with 
the numerical Bonn wave functions as well as J. Carbonell and 
V.A. Karmanov for communicating their own results.

\appendix

\section{Summation of all contributions to the interaction at the zeroth order 
in $\frac{1}{m}$}

We here discuss the summation of a subset of diagrams contributing to the two 
body interaction in the non-relativistic limit. Our intent is to show that the 
cancellation of second order contributions arising mainly from renormalization 
and crossed boson exchange has a somewhat general character for neutral spinless 
bosons, leaving the single boson exchange contribution as the only effective 
one. Before entering into the details, let's mention that this result has been 
extended to three bosons. Beside all irreducible boson exchange diagrams, it 
supposes to take into account corrections of the order $V_1^2$ in the 
renormalized interaction, Eq. (13), as well as second order contributions in the 
expansion of the energy denominators given by Eq. (3).

The developments follow those presented for instance in ref. \cite{DESP3}, 
but with a difference as to eliminating the elementary interaction of the 
constituents with the bosons. This one will be carried only on the part of 
the Hamiltonian containing the boson mass term, $ H_0(b)$, and the interaction 
term, $ H_I(c,b) $. The residual interaction of constituents with the bosons, 
generated by the transformation of the kinetic Hamiltonian of the constituents, 
$ H_0(c) $, will be of the first order in the inverse of their mass, implying 
second order contributions for the interaction between the constituents, hence 
the announced result. 

The schematic Hamiltonian we start from may be written as:
\begin{equation}
H(c,b)= H_0(c) + H_0(b) + H_I(c,b), 
\end{equation}
with
\begin{displaymath}
H_0(c) = \sum_{\vec{p}} E_{p} \, c^{+}_{\vec{p}} c_{\vec{p}}, \;\;
H_0(b)= \sum_{\vec{k}} \omega_{k} \, b^{+}_{\vec{k}} b_{\vec{k}}, \;\;
H_I(c,b)= g \sum_{\vec{p},\vec{k}} c^{+}_{\vec{p}+\vec{k}} c_{\vec{p}} 
\; \frac{(b^{+}_{-\vec{k}}+b_{\vec{k}})}{\sqrt{2\omega_{k}}},
\end{displaymath}
where the destruction operators, $c_{\vec{p}} $ and $b_{\vec{k}}$, (and the 
cooresponding creation oprators), respectively refer to the 
constituents and the bosons, whose exchange is responsible for the 
interaction between constituents. 

We now make the substitution
\begin{equation}
  [c^{+}, c \;, b^{+}, b] = e^{S(C^{+},C\;,B^{+},B)} \; [C^{+},C\;,B^{+},B] \; 
   e^{-S(C^{+},C\;,B^{+},B)},
\end{equation}
where $ S(C^{+},C\;,B^{+},B) $ is an antihermitic operator given by :

\begin{equation}
  S(C^{+},C\;,B^{+},B)=\sum_{\vec{p},\vec{k}} \frac{g}{ \sqrt{ 2\omega_{k} } } 
C^{+}_{\vec{p}+\vec{k}}   C_{\vec{p}} \; 
\left( \frac{B^{+}_{-\vec{k}}}{\omega_{k}} 
    -  \frac{B_{\vec{k}}}{\omega_{k}}      \right) .
\end{equation}
The above transformation, which allows one to express the bare degrees of 
freedom (c, b) in terms of the effective ones (C, B), is a unitary one, leaving 
commutation relations unchanged. It is no more than a change of basis in the 
Fock space. With the above substitution, the Hamiltonian
is written:
\begin{equation}
  H(c,b) = e^S H(C,B) \, e^{-S}= H(C,B)+ [S, H(C,B)] 
  + \frac{1}{2} \, [S,[S,H(C,B)]] +...
\end{equation}
Concentrating on that part involving the boson mass and interaction 
terms in $ H(c,b) $, one can show that the choice made in Eq. (58) allows 
one to remove the interaction term linear in the coupling g. For this 
purpose, one uses the relation: 
\begin{equation}
 H_I(C,B)+ [S, H_0(B)]=0.
\end{equation}
Taking into account this relation, the next contribution of the second 
order in the coupling g involves the double commutator, $ [S,[S,H_0(B)]] $. 
There are two terms arising from the second and third terms in Eq. (59), 
with the factors $-1$ and  $\frac{1}{2}$ respectively. The resulting 
contribution is nothing but the usual two-body interaction between the 
constituents:
\begin{equation}
 H_{2body}(C)=-\frac{1}{2} \sum_{\vec{p},\vec{p}',\vec{k}} 
 \frac{g^2}{\omega_{k}^2 } \; 
C^{+}_{\vec{p}+\vec{k}}  C_{\vec{p}}  \; C^{+}_{\vec{p}'-\vec{k}} C_{\vec{p}'}.
\end{equation}
Beyond the double commutator, it can be checked that the multiple commutators of 
$ S $ with $ H_0(B) $ vanish, leaving the two-body interaction, Eq. (61), as 
the 
only contribution from the boson mass and interaction terms of the original 
Hamiltonian. It is identical to the single boson exchange potential in the 
instantaneous approximation. Notice that to obtain this result, $ S $ has to 
commute with $H_{2body}(C)$, which is only fulfilled for spin- and charge-less 
bosons. 
While the above contribution is quite simple, the contribution arising from the 
mass term of the constituents, $ H_0(c) $, is not. At the lowest order in the 
coupling g, it produces an interaction term given by:
\begin{equation}
\tilde{H}_I(C,B)= \sum_{\vec{p},\vec{k}} \frac{g}{\sqrt{2\omega_{k}}} 
C^{+}_{\vec{p}+\vec{k}} C_{\vec{p}} 
 \left(B^{+}_{-\vec{k}} - B_{\vec{k}} \right) \; 
\left( \frac{ E_{\vec{p}}-E_{\vec{p}+\vec{k}} }{ \omega_{k} } \right)  .
\end{equation}
As seen from the above expression, the effect the interaction may produce is of 
the order of the difference in the energies of the constituents, namely 
$\frac{p^2}{m }$,  which means an effect of the order $(\frac{p^2}{m 
})^2$ for the interaction of the constituents. This is a typical 
relativistic correction of the order $(\frac{v}{c})^2$, expected in any case. As 
shown in the text of the paper in a limited case, the various corrections to the 
effective two-body interaction beyond the instantaneous approximation therefore 
cancel out at the zeroth order in $\frac{1}{m}$ for spin- and charge-less 
bosons.

\normalsize



\begin{thebibliography}{99} 
\bibitem{BETH} Salpeter, E.E. and Bethe, H.A.: Phys. Rev. {\bf 84}, 1232 (1951).
\bibitem{FRIA1} Friar, J.L.: Phys. Rev. {\bf C22}, 796, (1980).
\bibitem{ITZY} Itzykson, C. and Zuber, J.B. (eds.): {\em Quantum Field Theory}. 
McGraw-Hill  International  Editions (1985).
\bibitem{WICK} Wick, G.C.: Phys. Rev. {\bf 96}, 1124 (1954).
\bibitem{CUTK} Cutkosky, R.E.: Phys. Rev. {\bf 96}, 1135 (1954).
\bibitem{FELD} Feldman, G. , Fulton, T. and Townsend, J.: Phys. Rev. {\bf D7}, 
  1814 (1973).
\bibitem{FRED} Frederico, T. and Schulze, R.W.: Phys. Rev. {\bf C54}, 2201 
  (1996).
\bibitem{DESP2} Gaspard, P.: stage DEA, unpublished (1996).
\bibitem{CARB1} Mangin-Brinet, M.:  stage, unpublished (1997).
\bibitem{CARB2} Carbonell, J. and  Karmanov, V.A.: Nucl. Phys. {\bf A581}, 
  625 (1995).
\bibitem{PFEI} Pfeiffer, I.: stage, unpublished (1997).
\bibitem{BONN} Machleidt, R., Holinde, K. and Elster, Ch.: Phys. Rep. {\bf 149}, 
  1 (1987).
\bibitem{DESP1} Desplanques, B.: Phys. Lett. {\bf 203B}, 200 (1988). 
\bibitem{CARB3} Carbonell, J. et al. : Phys. Reports  {\bf 300}, 215  (1998). 
\bibitem{KARM} Karmanov, V.A.: Sov. J. Part. Nucl. {\bf 19}, 228 (1988). 
\bibitem{KEIS} Keister, B.D. and Polyzou, W.N.: Advances in Nuclear Physics 
  {\bf 20},  225 (1991).  
\bibitem{MACH} Machleidt, R.: Adv. in Nucl. Phys. {\bf 19}, 189 (1989).
\bibitem{FRIA2} Friar, J. : In: {\em  Nuclear Physics  with Electromagnetic 
  Probes} ( Lectures Notes in Physics 108; Arenh\"ovel, H. and Drechsel, D. 
(eds.) ).   Springer-Verlag (1979).
\bibitem{FST}  Fukuda, N., Sawada, K. and Taketani, M.:
  Prog. Theor. Phys. {\bf 12}, 156 (1954). 
\bibitem{OKUB} Okubo, S.: Prog. Theor. Phys. {\bf 12}, 603 (1954); \\Sugawara, 
M. and Okubo, S. : Phys. Rev. {\bf 117}, 605 (1960).
\bibitem{JOHN} Johnson, M.B.: Ann. Phys. {\bf 97}, 400 (1976).
\bibitem{SATO} Sato, T. et al.: J. Phys. {\bf G17}, 303 (1991).
\bibitem{AMGH} Amghar, A. and Desplanques, B.: Nucl.  Phys. {\bf A585}, 
  657 (1995).
\bibitem{CHEM} Chemtob, M.: In: {\em Mesons and Nuclei}, p. 496; 
  Rho, M. and Wilkinson, D.H. (eds.). North-Holland Publishing Company (1979).
\bibitem{GARI} Gari, M. and Hyuga, H.: Nucl. Phys.  {\bf A264}, 409 (1976).  
\bibitem{HYUG} Hyuga, H. and Gari, M.: Nucl.  Phys. {\bf A274}, 333 (1976).  
\bibitem{MAHA} Bortignon, P.F., Broglia, R.A., Dasso C.H. and Mahaux, C.:
  Phys. Reports {\bf 120}, 1 (1985).
\bibitem{NOGU} Noguera, S. and Desplanques, B.: 
  Phys. Lett. {\bf 149B}, 272 (1984). 
\bibitem{DESP3} Desplanques, B.: 
In: {\em Trends in Nuclear Physics, 100 years later}, les Houches LXVI (1996);  
  Nifenecker , H. et al. (eds.). North-Holland (1998).
\bibitem{PARI} Lacombe, M. et al.: Phys. Rev. {\bf C21}, 861 (1980). 
\bibitem{THEU} Desplanques, B., Theussl, L. and Amghar, A.: in preparation
\bibitem{SILV}  Silvestre-Brac, B. et al.: Phys. Rev. {\bf D29}, 2275 (1984).
\bibitem{BILA} Bilal, A. and Schuck, P.: Phys. Rev. {\bf D31}, 2045 (1985).
\bibitem{JI} Ji, C.R. and Furnstahl, R.J.: Phys. Lett. {\bf 167B}, 11 (1986).
\bibitem{KARM2} Karmanov, V.A. and Smirnov, A.: Nucl. Phys. 
  {\bf A575}, 520 (1994).
\bibitem{TJON} Zuilhof, M. and Tjon, J.: Phys. Rev. {\bf C24}, 736 (1981).
\bibitem{GROS} Arnold, R.G., Carlson, C.E. and Gross, F.: 
  Phys. Rev. {\bf C23}, 363 (1981).
\bibitem{DESP7} Desplanques, B. et al.: Z. Phys., Hadrons and Nuclei, 
  {\bf A343}, 331 (1992).
\bibitem{VALC} Fernandez, F., Gonz\'alez, A. and Valcarce, A.: 
Few-Body Systems, Suppl. {\bf 10}, 395 (1999).
\bibitem{BREZ} Brezin, E., Itzykson, C. and Zinn-Justin, J.: 
  Phys. Rev. {\bf D1}, 2349 (1970).
\bibitem{TODO} Todorov, I.T.: Phys. Rev. {\bf D3}, 2351 (1971).
\bibitem{GROS2} Gross, F.: Phys.Rev. {\bf C26}, 2203 (1982).
\bibitem{NEGH} Neghabian, A.R. and Gl\"ockle, W.: 
  Can. J. Phys. {\bf 61}, 85 (1983).
\bibitem{NIEU} Taco Nieuwenhuis and Tjon, J.A.: Phys. Rev. Lett. 
  {\bf 77}, 814 (1996).
\bibitem{DARE} Darewych, J.W.: Can. J. Phys. {\bf 76}, 523 (1998).
\bibitem{PAUS} Pauschenwein, J., Mathelitsch, L. and Plessas, W.: 
  Nucl. Phys. {\bf A508}, 253c (1990).
\bibitem{AMGH1} Amghar, A.: Th\`ese de Doctorat de l'Universit\'e Joseph
  Fourier, Grenoble I, unpublished (1993).  
\bibitem{DESP5} Desplanques, B.: Few-Body Systems, Suppl. {\bf 5}, 260 (1992).
\end{thebibliography}
\end{document}